\documentclass[prb,twocolumn,showpacs,aps,superscriptaddress,floatfix]{revtex4}
\usepackage{amsmath}
\usepackage{amssymb}
\usepackage{amsfonts}
\usepackage{bm}
\usepackage{graphicx}
\usepackage{color}
\usepackage[svgnames]{xcolor}
\usepackage{ulem}
\usepackage{hyperref}
\usepackage{verbatim}

\DeclareMathOperator{\Real}{Re}
\DeclareMathOperator{\Imag}{Im}

\begin{document}

\title{Charge distribution and spin textures in magic-angle twisted bilayer graphene}

\author{A.O. Sboychakov}
\affiliation{Institute for Theoretical and Applied Electrodynamics, Russian Academy of Sciences, Moscow, 125412 Russia}

\author{A.V. Rozhkov}
\affiliation{Institute for Theoretical and Applied Electrodynamics, Russian Academy of Sciences, Moscow, 125412 Russia}

\author{A.L. Rakhmanov}
\affiliation{Institute for Theoretical and Applied Electrodynamics, Russian Academy of Sciences, Moscow, 125412 Russia}

\begin{abstract}
We examine the coexisting spin and charge density waves as a possible ground state of the magic-angle twisted bilayer graphene.  When interactions are not included, the spectrum of the material has 4 (8 if spin is taken into account) almost flat almost degenerate bands. Interactions break down the degeneracy forming an order parameter which is usually assumed to be a spin density wave with a preset spin structure. Here we take into account a possible charge density wave contribution to the order parameter, that is, inhomogeneous distribution of the charge density within a twisted graphene supercell. We also calculate self-consistently the spin structure of the order parameter. We find that the density wave order is stable in the whole doping range from $-4$ to $+4$ extra electrons per supercell. The spin texture changes from collinear at zero doping to almost coplanar at finite doping. The density wave order shows nematic distortion when we dope the system. We demonstrate that the local spin magnetization is much stronger than the charge density variation, unless the doping exceeds $3$ extra electrons or holes per supercell.
\end{abstract}


\date{\today}

\maketitle

After the discovery of Mott insulating states~\cite{NatureMott2018,MottSCNature2019} and superconductivity~\cite{MottSCNature2019,NatureSC2018} in twisted bilayer graphene (tBLG), this material is intensively studied. In tBLG one graphene layer is rotated with respect to another one by a twist angle $\theta$. As a result, the system has a superstructure for certain commensurate twist angles~\cite{ourBLGreview2016}. The low-energy electronic properties substantially depend on $\theta$. At the so-called first magic angle $\theta_c\cong1^{\circ}$ the energy spectrum  is characterized by 4 (8 if spin is included) almost flat bands close to the Fermi level. This flatness makes the system very susceptible to electron-electron interactions. Interactions lead to the spontaneous violation of certain symmetries. The nature of non-superconducting many-body states in tBLG is not yet definitively known. Different candidates were proposed for the observed many-body insulating states in the tBLG~\cite{Philips2018,PhysRevB.98.081102,ChiralSDW_SC2018,
AFMMottSC2019,FMorderPRL2019,OurtBLGPRB2019,NematicPRB2020,
PhysRevLett.128.247402,PhysRevLett.128.156401,JETPLetters.112.651}.
The spin density wave (SDW) is among them~\cite{ChiralSDW_SC2018,AFMMottSC2019,OurtBLGPRB2019,NematicPRB2020}.

In our previous papers~\cite{OurtBLGPRB2019,NematicPRB2020} we assumed the
planar SDW order to be the ground state of the system in the doping range
$-4<x<4$ electrons per supercell. This choice was not arbitrary. It is
known that in the tBLG at small twist angle the electrons at the Fermi
level are located in the AA regions of the superlattice
cell~\cite{ourBLGreview2016,dSPRB}. The ground state of the AA stacked
bilayer graphene should be antiferromagnetic
(AFM)~\cite{AAPRL,AAPRB2013}.
Based on the assumption of the SDW order, our simulations reproduced
qualitatively well the dependence of the conductivity on the doping
observed experimentally. We also
showed~\cite{NematicPRB2020}
that doping the system away from the charge neutrality point leads to the
formation of the nematic state, which is also detected by the
STM~\cite{MottNematicNature2019,KerelskyNematicNature2019,STMNature2019,WongNematic2020}.

In the present paper we extend the variety of possible order parameters, considering the non-coplanar SDW order on a backdrop of the inhomogeneous charge distribution. We show that the spin texture changes from collinear to almost coplanar with doping. It is demonstrated that the symmetry of the SDW order is reduced with doping. This leads to the formation of the nematic spin state. At the same time, the rotational symmetry of the charge density is almost unaffected by doping. The nematic state can manifests itself in the non-symmetric spatial distribution of the local density of states, which was observed experimentally~\cite{MottNematicNature2019,KerelskyNematicNature2019,STMNature2019,WongNematic2020}. Finally, we show that local spin density is much stronger than the charge density variation unless the doping exceeds 3 extra electrons or holes per supercell.

\textbf{Model Hamiltonian.}
The periodic superstructure exists in the tBLG when $\theta$ satisfies the condition
\begin{eqnarray}
\cos\theta=(3m_0^2+3m_0r+r^2/2)/(3m_0^2+3m_0r+r^2),
\end{eqnarray}
with $m_0$ and $r$ being mutually coprime positive integers~\cite{ourBLGreview2016}. The superlattice cell has a form of a right rhombus. When $r=1$ the superlattice cell coincides with the moir{\'{e}} cell. In our study we consider only such superstructures. When the twist angle is small, the supercell can be considered as consisting of regions with AA, AB, and BA stacking~\cite{ourBLGreview2016,NanoLettTB}.

We use the following model Hamiltonian of tBLG:
\begin{eqnarray}
\label{H}
H&=&\!\!\!\sum_{{\mathbf{nm}ij\atop\alpha\beta\sigma}}\!t(\mathbf{r}_{\mathbf{n}}^{i\alpha};\mathbf{r}_{\mathbf{m}}^{j\beta})d^{\dag}_{\mathbf{n}i\alpha\sigma}d^{\phantom{\dag}}_{\mathbf{m}j\beta\sigma}+
U\!\sum_{{\mathbf{n}i\alpha\sigma}}\!n_{\mathbf{n}i\alpha\uparrow}n_{\mathbf{n}i\alpha\downarrow}+\nonumber\\
&&\frac12\!\mathop{{\sum}'}_{{\mathbf{nm}ij\atop\alpha\beta\sigma\sigma'}}\!V(\mathbf{r}_{\mathbf{n}}^{i\alpha}-\mathbf{r}_{\mathbf{m}}^{j\beta})n_{\mathbf{n}i\alpha\sigma}n_{\mathbf{m}j\beta\sigma'}\,.
\end{eqnarray}
Here $d^{\dag}_{\mathbf{n}i\alpha\sigma}$ ($d^{\phantom{\dag}}_{\mathbf{n}i\alpha\sigma}$) are the creation (annihilation) operators of the electron with spin $\sigma$\,($=\uparrow$, $\downarrow$) at the unit cell $\mathbf{n}$ in the layer $i$\,($=1,2$) in the sublattice $\alpha$\,($={\cal A,B}$), while $n_{\mathbf{n}i\alpha\sigma}=d^{\dag}_{\mathbf{n}i\alpha\sigma}d^{\phantom{\dag}}_{\mathbf{n}i\alpha\sigma}$. The first term in Eq.~\eqref{H} is the single-particle tight-binding Hamiltonian with $t(\mathbf{r}_{\mathbf{n}}^{i\alpha};\mathbf{r}_{\mathbf{m}}^{j\alpha})$ being the amplitude of the electron hopping from site in the position $\mathbf{r}_{\mathbf{m}}^{j\beta}$ to the site $\mathbf{r}_{\mathbf{n}}^{i\alpha}$. The second term in Eq.~\eqref{H} describes the on-site (Hubbard) interaction of electrons, while the last term corresponds to the inter-site Coulomb interaction (the prime means that the elements with $\mathbf{r}_{\mathbf{n}}^{i\alpha}=\mathbf{r}_{\mathbf{m}}^{j\beta}$ should be excluded).

Now we have to choose a parametrization of the hopping amplitudes. We keep only nearest-neighbor terms for the intralayer hopping with $t=-2.57$\,eV. The inter-plane hopping amplitudes are parameterized by the following Slater-Koster formula for $p_z$ electrons
\begin{equation}
\label{ParI}
t(\mathbf{r};\mathbf{r}')=\frac{\left[(\mathbf{r}-\mathbf{r}')\mathbf{e}_z\right]^2}{|\mathbf{r}-\mathbf{r}'|^2}V_{\sigma}(\mathbf{r}-\mathbf{r}')\,,
\end{equation}
where $\mathbf{e}_z$ is the unit vector transverse to the layers,
\begin{equation}
\label{Vsigma}
V_{\sigma}(\mathbf{r})=t_0e^{-(|\mathbf{r}|-d)/r_0} F_c(|\mathbf{r}|)\,,\;\;F_c(r)=\frac{1}{1+e^{(r-r_c)/l_c}}\,,
\end{equation}
and the cutoff function $F_c(r)$ is introduced to nullify the hopping amplitudes at distances larger than $r_c$. We use $r_c=4.92$\,\AA, $l_c=0.2$\,\AA. The parameter $t_0$ defines the largest interlayer hopping  amplitude. We choose $t_0=0.37$\,eV (this value was used to describe the AB bilayer graphene~\cite{ourBLGreview2016}). The parameter $r_0$ describes how fast the hopping amplitudes decay inside the region $r<r_c$. We choose $r_0=0.34$\,\AA.

At the first magic angle $\theta=\theta_c$ the low-energy band structure
consists of 4 almost flat almost degenerate bands separated by energy gaps
from lower and higher dispersive bands. The bandwidth $W$ of these flat
bands has a minimum at $\theta=\theta_c$. For model parameters used here
we numerically determine that
$W=1.8$\,meV, the gaps between flat and dispersive bands are  $\approx 2.5$\,meV, and $\theta_c=1.08^\circ$ (which is close to the experimentally observed and corresponds to the superstructure with $m_0=30$ and $r=1$).

\textbf{Spin and Charge Densities.}
When the system has flat bands crossing the Fermi level, the interactions
become very important. The interactions break the symmetry of the single
particle Hamiltonian inducing some order parameter. We start with the
SDW-like ordering, with multicomponent order parameter. First, we introduce on-site order parameters
\begin{eqnarray}
\label{Deltania}
\Delta_{\mathbf{n}i\alpha}&=&U\langle d^{\dag}_{\mathbf{n}i\alpha\uparrow}d^{\phantom{\dag}}_{\mathbf{n}i\alpha\downarrow}\rangle\,,\nonumber\\
\Delta^{z}_{\mathbf{n}i\alpha}&=&U\left[\langle d^{\dag}_{\mathbf{n}i\alpha\uparrow}d^{\phantom{\dag}}_{\mathbf{n}i\alpha\uparrow}\rangle-
\langle d^{\dag}_{\mathbf{n}i\alpha\downarrow}d^{\phantom{\dag}}_{\mathbf{n}i\alpha\downarrow}\rangle\right]\!/2.
\end{eqnarray}
The quantity $\Delta_{\mathbf{n}i\alpha}$ is complex, while $\Delta^{z}_{\mathbf{n}i\alpha}$ is real. They describe on-site magnetization
\begin{equation}
\label{S}
\mathbf{S}_{\mathbf{n}i\alpha}=\left[\Real(\Delta_{\mathbf{n}i\alpha}),\,\Imag(\Delta_{\mathbf{n}i\alpha}),\,\Delta^{z}_{\mathbf{n}i\alpha}\right]/U.
\end{equation}
Parameters $\Delta_{\mathbf{n}i\alpha}$ and $\Delta^{z}_{\mathbf{n}i\alpha}$ are controlled by the Hubbard interaction. We take $U=2t$. This value is somewhat smaller than the critical value for a single-layer graphene transition into a mean-field AFM state~\cite{MF_Uc_sorella1992}, $U_c=2.23t$. Thus, our Hubbard interaction is rather strong, but not too strong to open a gap in single layer graphene.

In a graphene layer, each atom in one sublattice has three nearest neighbors belonging to another sublattice. For this reason we consider three types of in-plane nearest-neighbor order parameters of the SDW type, $A^{(\ell)}_{\mathbf{n}i\sigma}$ and $A^{z(\ell)}_{\mathbf{n}i}$ ($\ell=1,\,2,\,3$), corresponding to three different links connecting the nearest-neighbor sites. These order parameters are defined as follows
\begin{eqnarray}
\label{Anis}
A^{(\ell)}_{\mathbf{n}i\sigma}&=&V_{\rm nn}\langle	d^{\dag}_{\mathbf{n}+\mathbf{n}_{\ell}i{\cal A}\sigma}	d^{\phantom{\dag}}_{\mathbf{n}i{\cal B}\bar{\sigma}}\rangle\,,\\
A^{z(\ell)}_{\mathbf{n}i}&=&\frac{V_{\rm nn}}{2}\left(\langle	 d^{\dag}_{\mathbf{n}+\mathbf{n}_{\ell}i{\cal A}\uparrow}	d^{\phantom{\dag}}_{\mathbf{n}i{\cal B}\uparrow}\rangle-
\langle	d^{\dag}_{\mathbf{n}+\mathbf{n}_{\ell}i{\cal A}\downarrow}	 d^{\phantom{\dag}}_{\mathbf{n}i{\cal B}\downarrow}\rangle\right),\nonumber
\end{eqnarray}
where
$\mathbf{n}_{1}=(0,\,0)$,
$\mathbf{n}_{2}=(1,\,0)$,
$\mathbf{n}_{3}=(0,\,1)$,
$\bar{\sigma}=-\sigma$,
and
$V_{\rm nn}=V
(|\bm{\delta}|)$
is the in-plane nearest-neighbor Coulomb repulsion energy
($\bm{\delta}$
is the vector connecting
${\cal A}$
and
${\cal B}$
sites in the graphene unit cell). We take
$V_{\text{nn}}/U=0.59$,
in agreement with Ref.~\cite{Wehling}. The order parameters~\eqref{Anis} define spins on links connecting in-plane nearest neighbor sites according to
\begin{eqnarray}
\label{Slink}
&&\!\!\mathbf{S}^{(\ell)}_{\mathbf{n}i}=\frac12\sum_{\sigma\sigma'}\!\bm{\sigma}_{\sigma\sigma'}\langle	 d^{\dag}_{\mathbf{n}+\mathbf{n}_{\ell}i{\cal A}\sigma}		 d^{\phantom{\dag}}_{\mathbf{n}i{\cal B}\sigma'}\rangle+{\rm c.c.}\\
&&\!\!=\frac{1}{V_{\text{nn}}}\!\left(\frac12\Real[A^{(\ell)}_{\mathbf{n}i\uparrow}+A^{(\ell)}_{\mathbf{n}i\downarrow}],\,\frac12\Imag[A^{(\ell)}_{\mathbf{n}i\uparrow}-	 A^{(\ell)}_{\mathbf{n}i\downarrow}],
A^{z(\ell)}_{\mathbf{n}i}\right),\nonumber
\end{eqnarray}
where $\bm{\sigma}$ is a vector of the Pauli matrices. In contrast to Ref.~\cite{NematicPRB2020}, the spins $\mathbf{S}_{\mathbf{n}i\alpha}$ and $\mathbf{S}^{(\ell)}_{\mathbf{n}i}$ are allowed to have the $z$ components. In Ref.~\cite{NematicPRB2020} we also considered
inter-layer nearest-neighbor SDW order parameters. The calculations showed, however, that these components are by order of magnitude smaller than $A^{(\ell)}_{\mathbf{n}i\sigma}$ (which are smaller than
$\Delta_{\mathbf{n}i\alpha}$). In present paper we neglect such order parameters.

Besides the SDW order parameters, we consider here the charge-density-wave-like contributions. First, we take into account that the charges are not uniformly distributed inside the superlattice cell and introduce the quantity
\begin{equation}
\label{Deltacnia}
\Delta^{c}_{\mathbf{n}i\alpha}=\frac{U}{2}\left[\sum_{\sigma}\langle d^{\dag}_{\mathbf{n}i\alpha\sigma}d^{\phantom{\dag}}_{\mathbf{n}i\alpha\sigma}\rangle-1\right].
\end{equation}
This value can be considered as the on-site potential due to charge inhomogeneity. Note that $\Delta^{c}_{\mathbf{n}i\alpha}$ is finite even in the absence of any symmetry breaking since the sites inside a supercell are non-identical. Thus, it cannot be considered as an order parameter. In our simulations $\Delta^{c}_{\mathbf{n}i\alpha}$ is normalized according to
\begin{equation}
\label{x}
\frac{2}{U{\cal N}_{sc}}\sum_{\mathbf{n}i\alpha}\Delta^{c}_{\mathbf{n}i\alpha}=x\,,
\end{equation}
where ${\cal N}_{sc}$ is the number of supercells in the system, and $x$ is the doping level, that is, the number of extra electrons or holes per one supercell. Besides $\Delta^{c}_{\mathbf{n}i\alpha}$, we introduce the in-plane nearest-neighbor (inter-site) potentials
\begin{equation}
\label{Acnis}
A^{c(\ell)}_{\mathbf{n}i}=\frac{V_{\rm nn}}{2}\sum_{\sigma}\langle	 d^{\dag}_{\mathbf{n}+\mathbf{n}_{\ell}i{\cal A}\sigma}	d^{\phantom{\dag}}_{\mathbf{n}i{\cal B}\sigma}\rangle\,.
\end{equation}
For all these charge and spin distributions we assume the same periodicity as the supercell periodicity.

\begin{figure*}[t]
\centering
\includegraphics[width=0.24\textwidth]{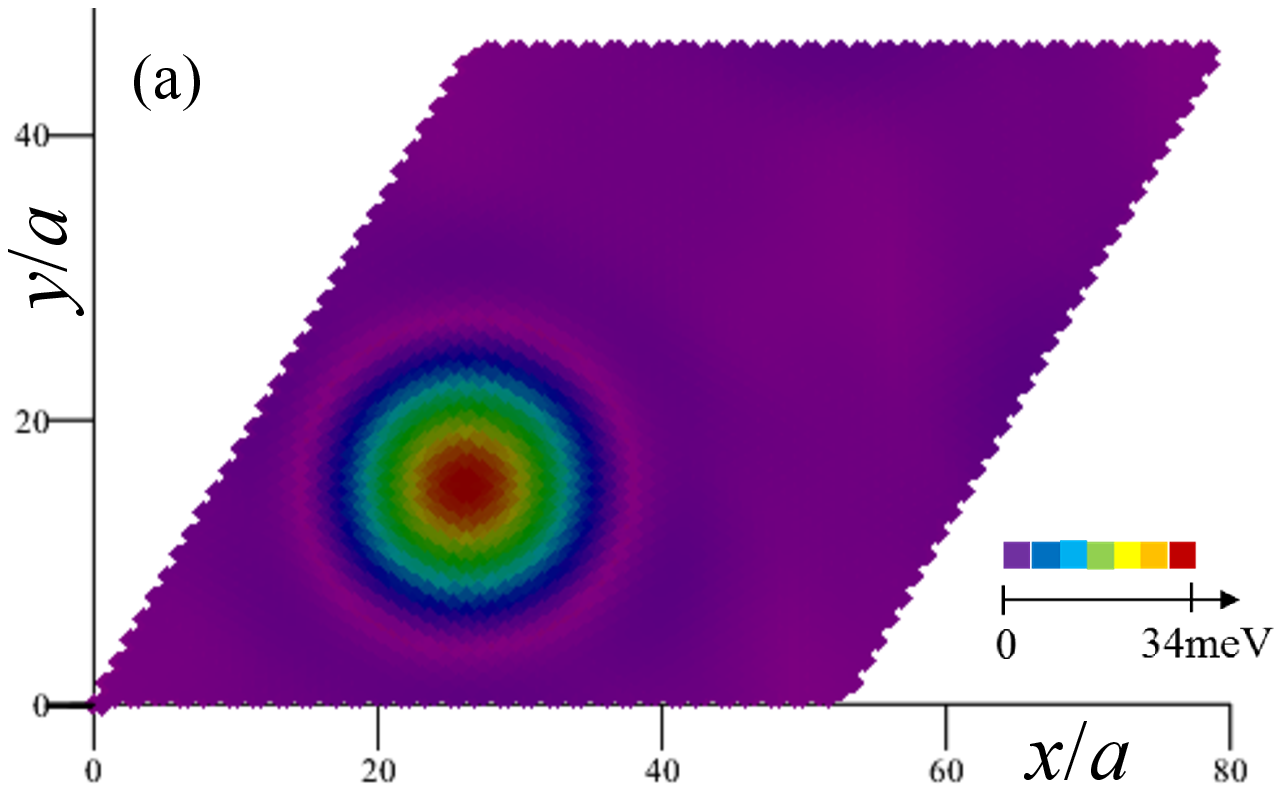}
\includegraphics[width=0.24\textwidth]{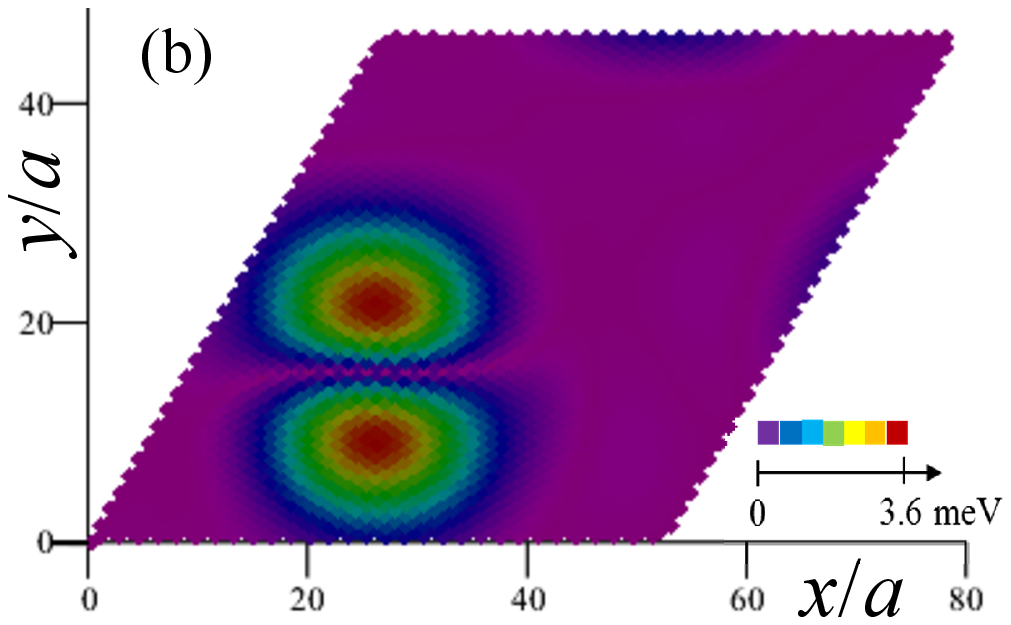}
\includegraphics[width=0.24\textwidth]{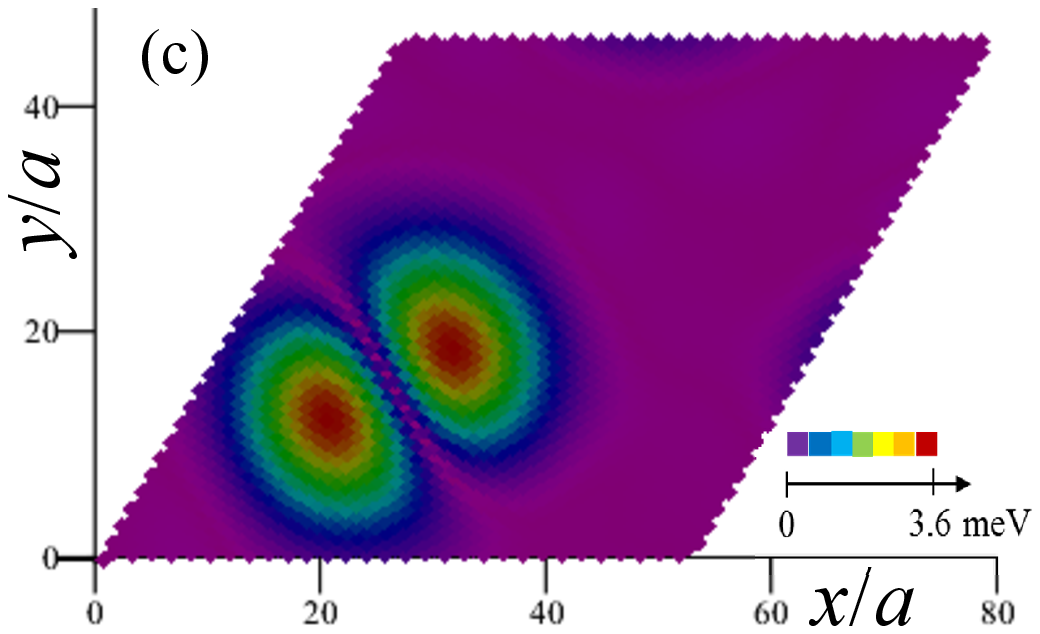}
\includegraphics[width=0.24\textwidth]{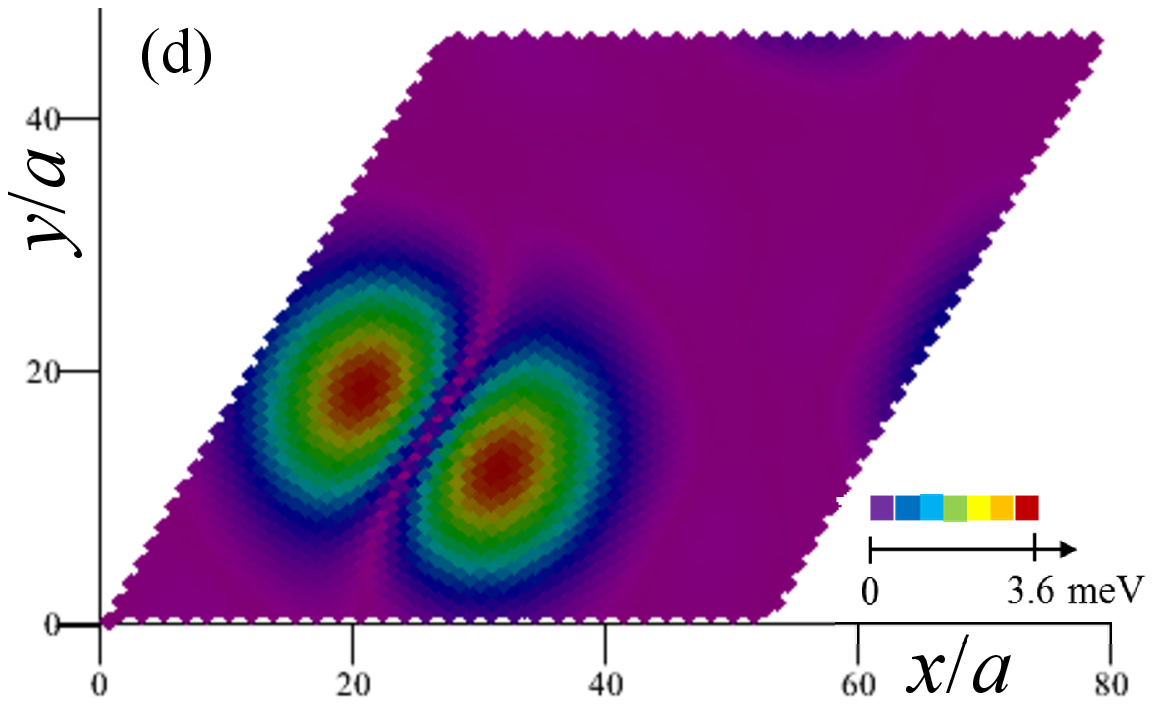}\\
\includegraphics[width=0.24\textwidth]{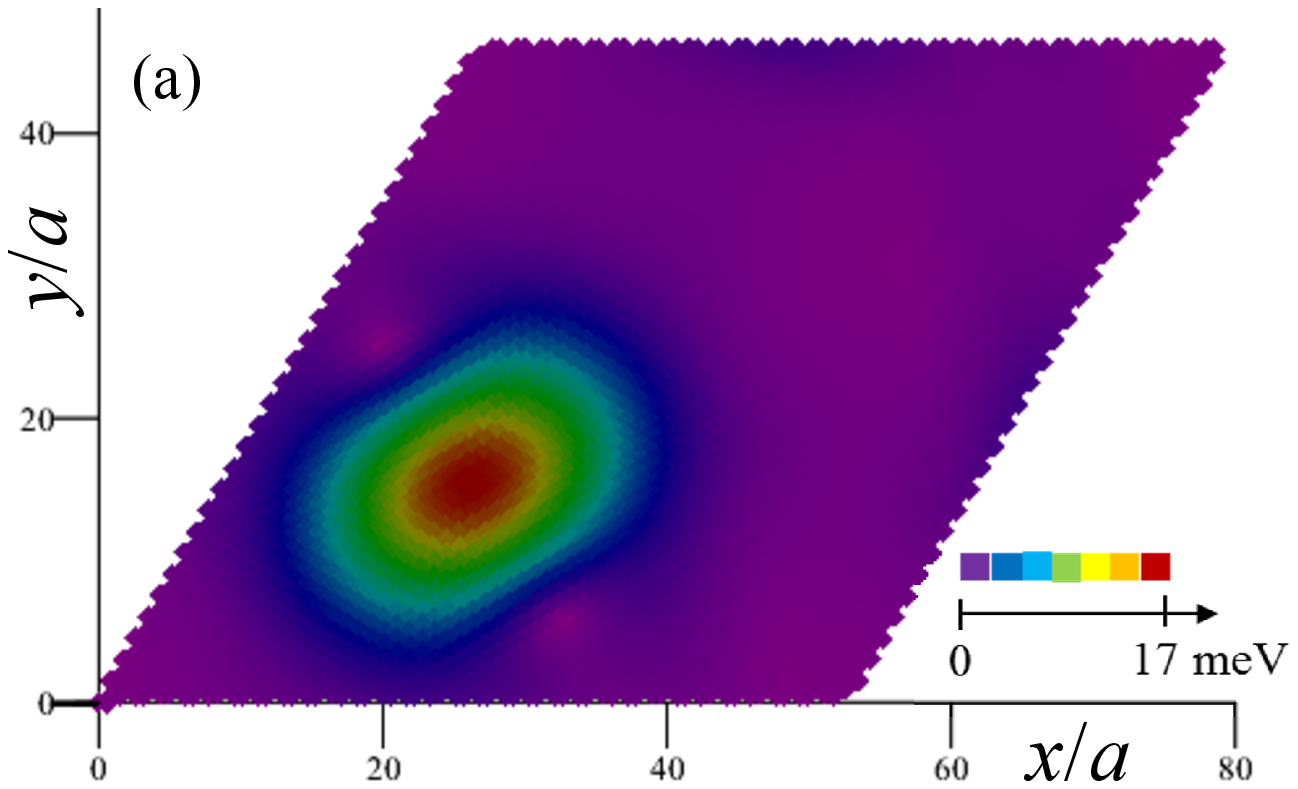}
\includegraphics[width=0.24\textwidth]{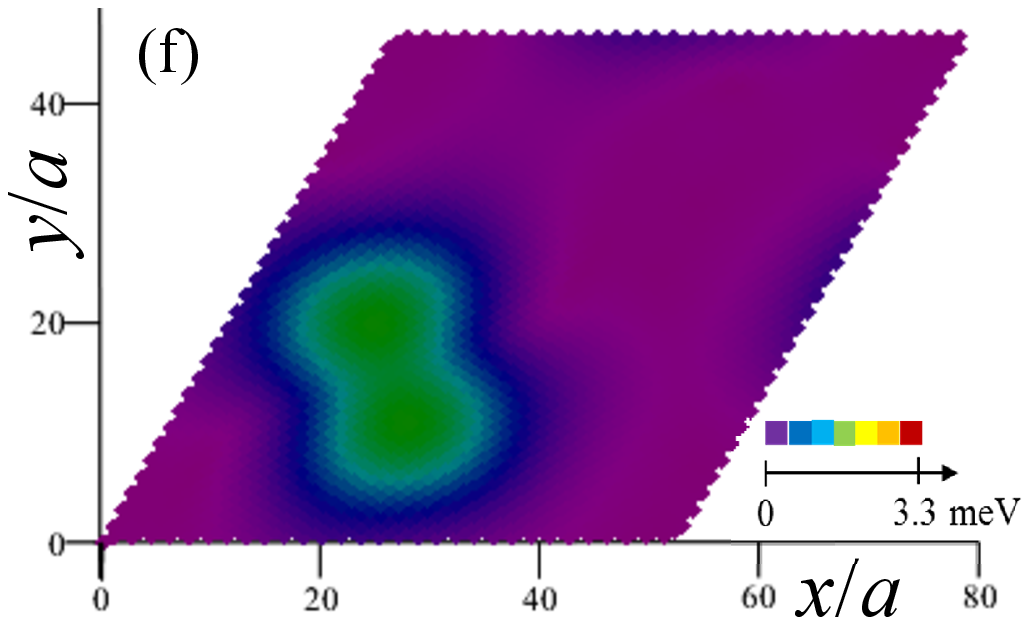}
\includegraphics[width=0.24\textwidth]{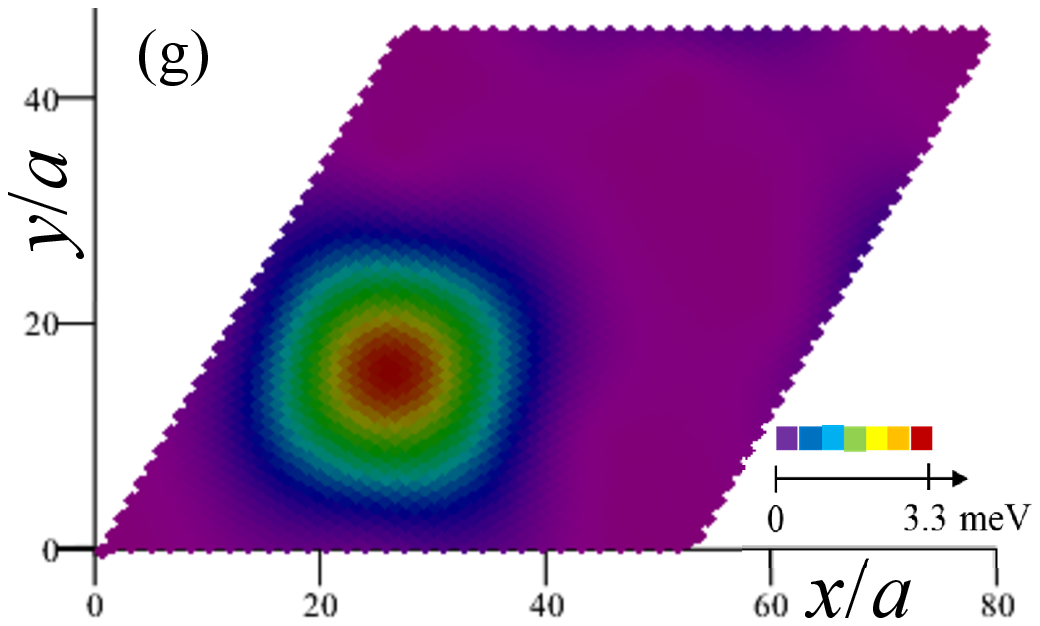}
\includegraphics[width=0.24\textwidth]{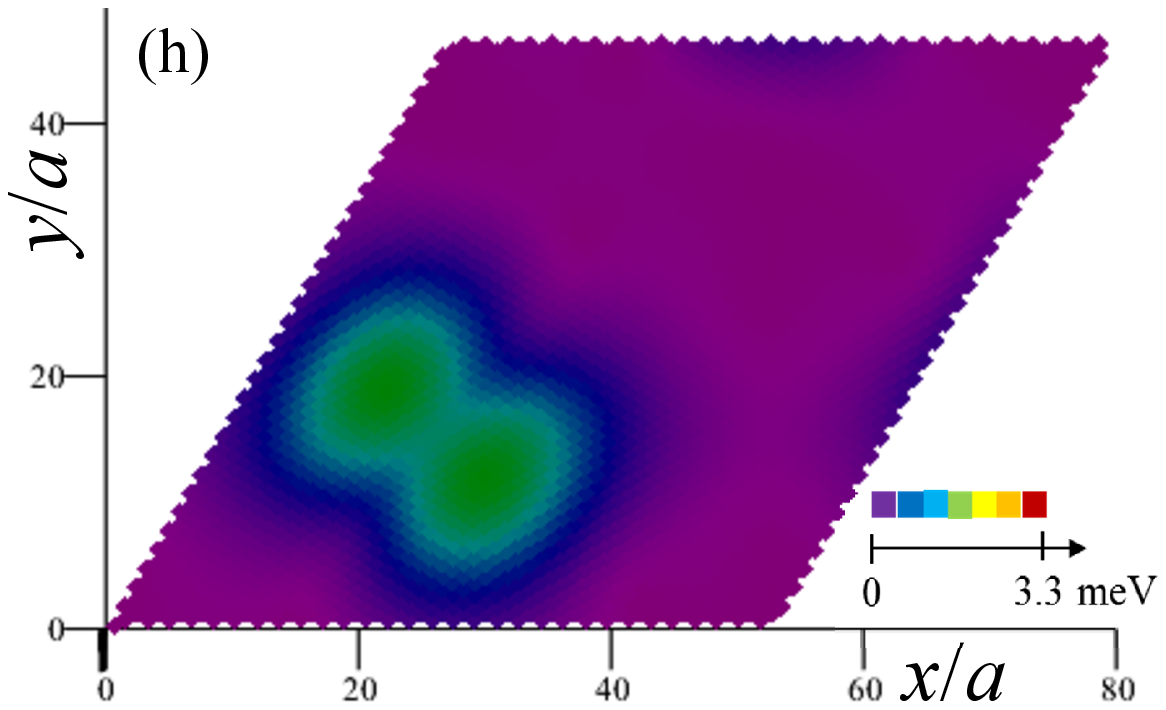}
\caption{\label{FigSDW}The spatial distribution of the order parameters ${\cal D}_{\mathbf{n}1\alpha}$ [(a), (e)] and ${\cal A}^{(\ell)}_{\mathbf{n}1}$ [ (b)\,--\,(d) and (f)\,--\,(h)] (for definition, see the text) calculated at $x=0$ [(a)\,--\,(d)] and $x=-2$ [(e)\,--\,(h)]. Doping reduces the symmetry of the order parameters from $C_6$ down to $C_2$.}
\end{figure*}

\textbf{Mean-field iteration scheme.}
To calculate the order parameters and potentials, we use a mean-field approach. It is based on the decoupling of the quadratic terms in the Hamiltonian~\eqref{H}:
\begin{eqnarray}
\label{decoupling}
&&\!\!\!\!\!\!n^{\phantom{\dag}}_{\mathbf{n}i\alpha\sigma}n^{\phantom{\dag}}_{\mathbf{m}j\beta\sigma'}\!\rightarrow\!
n^{\phantom{\dag}}_{\mathbf{n}i\alpha\sigma}\langle n^{\phantom{\dag}}_{\mathbf{m}j\beta\sigma'}\rangle
+n^{\phantom{\dag}}_{\mathbf{m}j\beta\sigma'}\langle n^{\phantom{\dag}}_{\mathbf{n}i\alpha\sigma}\rangle-\nonumber\\
&&\!\!\!\!\!\!\langle n^{\phantom{\dag}}_{\mathbf{n}i\alpha\sigma}\rangle\langle n^{\phantom{\dag}}_{\mathbf{m}j\beta\sigma'}\rangle
-d^{\dag}_{\mathbf{n}i\alpha\sigma}d^{\phantom{\dag}}_{\mathbf{m}j\beta\sigma'}\langle	 d^{\dag}_{\mathbf{m}j\beta\sigma'}	 d^{\phantom{\dag}}_{\mathbf{n}i\alpha\sigma}\rangle-\nonumber\\
&&\!\!\!\!\!\!d^{\dag}_{\mathbf{m}j\beta\sigma'}d^{\phantom{\dag}}_{\mathbf{n}i\alpha\sigma}\langle d^{\dag}_{\mathbf{n}i\alpha\sigma}d^{\phantom{\dag}}_{\mathbf{m}j\beta\sigma'}\rangle+\\
&&\!\!\!\!\!\! \langle d^{\dag}_{\mathbf{n}i\alpha\sigma}d^{\phantom{\dag}}_{\mathbf{m}j\beta\sigma'}\rangle\langle d^{\dag}_{\mathbf{m}j\beta\sigma'}d^{\phantom{\dag}}_{\mathbf{n}i\alpha\sigma}\rangle.
\nonumber
\end{eqnarray}
As a result, we obtain a mean-field Hamiltonian $H^{\text{MF}}$. Numerical algorithm for finding the order parameters and the potentials is similar to that used in Ref.~\cite{NematicPRB2020}. First, we write down the electronic operators in the momentum representation~\cite{PankratovPRB2013}
\begin{eqnarray}
d^{\phantom{\dag}}_{\mathbf{pG}i\alpha\sigma}=
\frac{1}{\sqrt{{\cal{N}}}}\sum_{\mathbf{n}}\exp{[-i(\mathbf{p}
+\mathbf{G})\mathbf{r}_{\mathbf{n}}^{i}]}d_{\mathbf{n}i\alpha\sigma}\,,
\end{eqnarray}
where ${\cal N}$ is the number of the graphene unit cells in one layer of the sample, $\mathbf{r}_{\mathbf{n}}^{i}$ is the position of the $\mathbf{n}$-th unit cell of the $i$-th layer, the momentum $\mathbf{p}$ lies in the first Brillouin zone of the superlattice, and $\mathbf{G}$ are the reciprocal vectors of the superlattice lying in the first Brillouin zone. The number of vectors $\mathbf{G}$ is equal to $N_{\rm sc}={\cal N}/{\cal N}_{sc}$ for each graphene layer.

The mean-field Hamiltonian $\hat{H}_{\mathbf{p}}$ is $N_R\times N_R$ matrix, where $N_R=8N_{\rm sc}$ (factor $8$ is due to the spin, layer, and sublattice indices). The rank of this matrix is too large to perform numerical integration over the quasimomentum $\mathbf{p}$ for a realistic time. Indeed, for the first magic angle $\theta_c=1.08^{\circ}$ we have $N_R=22328$. For this reason, we use simplifications. The main contribution to the order parameters comes from the low-energy states. Consequently, the contributions from other states can be approximated. In the limit of uncoupled ($t_0=0$) graphene layers and zero order parameters, the matrix $\hat{H}_{\mathbf{p}}$ is block-diagonal with the  $2\times2$ matrices on its diagonal
\begin{equation}
-t\left(\begin{array}{cc}
0&f_{\mathbf{p}+\mathbf{G}}^{i}\\
f_{\mathbf{p}+\mathbf{G}}^{i*}&0
\end{array}\right),
\end{equation}
where $f^{i}_{\mathbf{p}}=1+e^{-i\mathbf{pa}^{i}_1}+e^{-i\mathbf{pa}^{i}_2}$, $\mathbf{a}^{i}_{1,2}$ are the unit vectors of the $i$-th layer. The eigenenergies of such a matrix are $\pm t|f_{\mathbf{p}+\mathbf{G}}^{i}|$. The interlayer hopping amplitudes and order parameters are much smaller than $t$. As long as we are interested in low-energy features, we can use the truncated matrix $\hat{H}'_{\mathbf{p}}$ excluding the rows and columns in $\hat{H}_{\mathbf{p}}$ containing elements with $t|f_{\mathbf{G+p}}^{i}|>E_R$, where $E_R$ is the cutoff energy. The rank of the truncated matrix is $N'_R<N_R$. The eigenenergies $E^{(S)}_{\mathbf{p}}$ of $\hat{H}'_{\mathbf{p}}$, which lie close to $\pm E_R$, are calculated with significant errors. To fix this problem, we take into account only bands with $|E^{(S)}_{\mathbf{0}}|<E_0$, where $E_0<E_R$. The number of such bands is $N_0<N'_R$. We use $E_0=0.2t$, $E_R=0.4t$ ($N_0=480$, $N'_R=720$). Calculations with smaller and larger $E_R$ and $E_0$ show that the results are almost independent of these quantities.

The contribution to the total energy from the discarded states $E^{(S)}_{\mathbf{p}}<-E_0$
must be accounted for separately. Since $E_0$ is much larger than order parameters, this can be done perturbatively. The leading corrections to the total energy are quadratic in SDW order parameters $\Delta_{\mathbf{n}i\alpha}$,
$\Delta^{z}_{\mathbf{n}i\alpha}$,
$A^{(\ell)}_{\mathbf{n}i\sigma}$,
and
$A^{z(\ell)}_{\mathbf{n}i}$.
We assume that the proportionality coefficients are identical for all order parameters and are equal to
\begin{equation}
\label{eq::susceptibility_approx}
-\frac{1}{V_c(E_0)}=-\frac12\int_{E_0}^{3t}\!\!\!\!dE\,\frac{\rho_0(E)}{E}\,,
\end{equation}
where
$\rho_0(E)$
is the single-layer graphene density of states. Such a correction can be taken into account by the replacement in the mean-field Hamiltonian
\begin{eqnarray}
\label{Hrepl}
H^{\text{MF}}&\rightarrow&H'^{\text{MF}}-\sum_{\mathbf{n}i\alpha}\frac{|\Delta_{\mathbf{n}i\alpha}|^2}{V_c(E_0)}-\sum_{\mathbf{n}i\alpha}\frac{|\Delta^{z}_{\mathbf{n}i\alpha}|^2}{V_c(E_0)}\nonumber\\
&&-\sum_{\mathbf{n}i{\ell}\sigma}\frac{|A^{(\ell)}_{\mathbf{n}i\sigma}|^2}{V_c(E_0)}-\sum_{\mathbf{n}i{z\ell}\sigma}\frac{|A^{z(\ell)}_{\mathbf{n}i}|^2}{V_c(E_0)},
\end{eqnarray}
where $H'^{\text{MF}}$ is the Hamiltonian in the truncated basis. The contributions to $\Delta^c$ and $A^{c(\ell)}$ from the bands with $E^{(S)}_{\mathbf{p}}<-E_0$ are almost independent of the interactions and ordering. In the limit of uncoupled layers and zero order parameters, the contribution to $\Delta^{c}_{\mathbf{n}i\alpha}$ from each excluded band equals $U/(4N_{sc})$. As a result, we obtain the self-consistency condition for $\Delta^{c}_{\mathbf{n}i\alpha}$
\begin{eqnarray}
\label{Deltacnia_}
\frac{2}{U}\Delta^{c}_{\mathbf{n}i\alpha}&=&\frac{1}{N_{sc}}\!\mathop{{\sum}'}_{S}\!\!\sum_{\mathbf{GG}'\sigma}\!\!\int\!\frac{d^2\mathbf{p}}{v_{RBZ}}
\Phi^{(S)*}_{\mathbf{pG}i\alpha\sigma}\Phi^{(S)}_{\mathbf{pG}'i\alpha\sigma}\times\nonumber\\
&&e^{-i(\mathbf{G}-\mathbf{G}')\mathbf{r}_{\mathbf{n}}^{i}}\Theta\left(\mu-E^{(S)}_{\mathbf{p}}\right)-N_0/8N_{sc},
\end{eqnarray}
where $\Phi^{(S)}_{\mathbf{pG}i\alpha\sigma}$ are eigenfunctions of
$\hat{H}'_{\mathbf{p}}$, $\Theta(x)$ is the step function, $v_{RBZ}$ is the
area of the supercell's Brillouin zone, and $\mu$ is the chemical
potential. The summation in Eq.~\eqref{Deltacnia_} is performed over $N_0$
bands and truncated basis, and the equation for chemical potential is
\begin{eqnarray}
\label{x_}
\mathop{{\sum}'}_{S}\!\!\int\!\frac{d^2\mathbf{p}}{v_{RBZ}}\Theta\left(\mu-E^{(S)}_{\mathbf{p}}\right)-\frac{N_0}{2}=x\,.
\end{eqnarray}

\begin{figure}[t]
\centering
\includegraphics[width=0.23\textwidth]{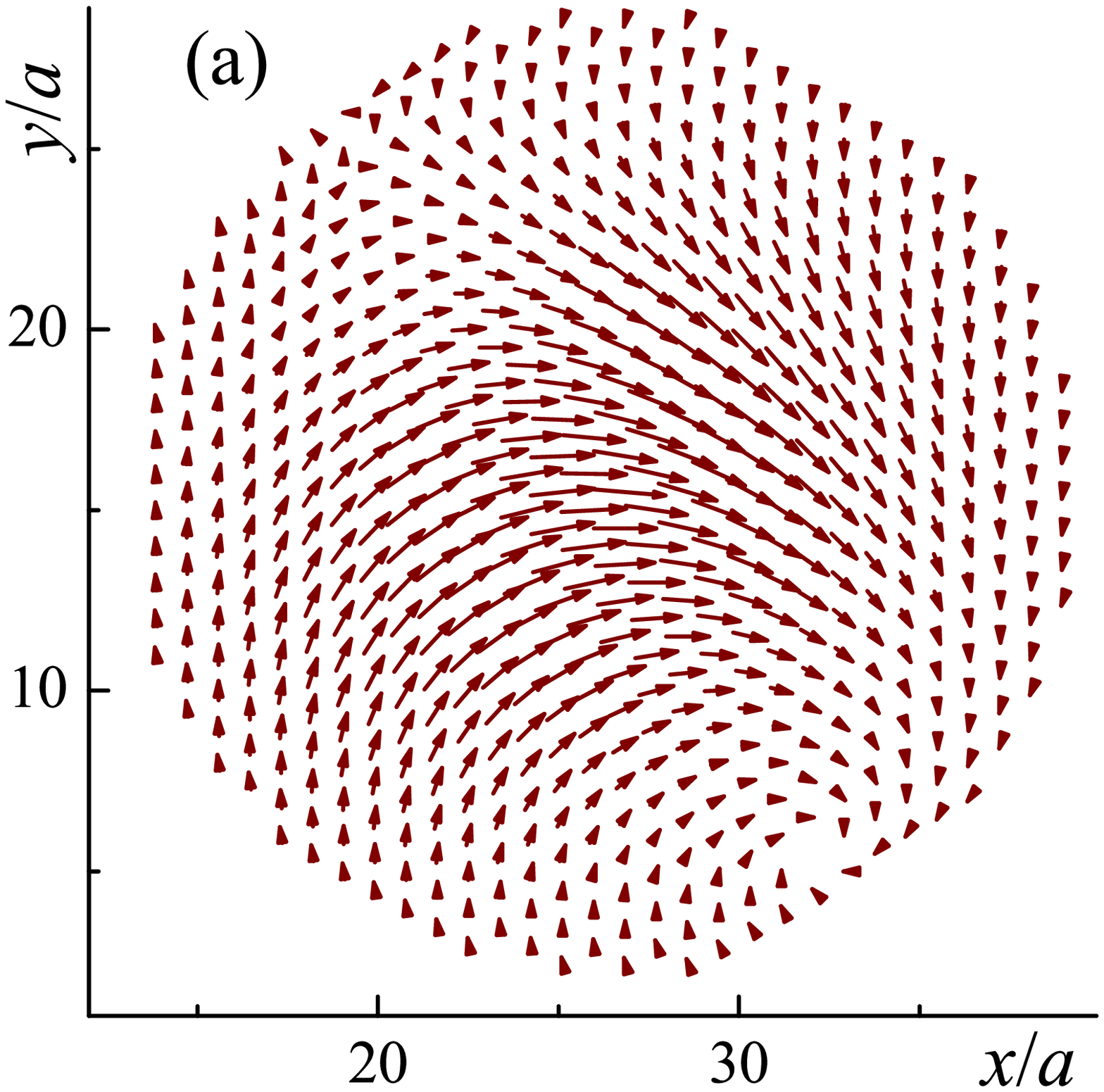}
\includegraphics[width=0.23\textwidth]{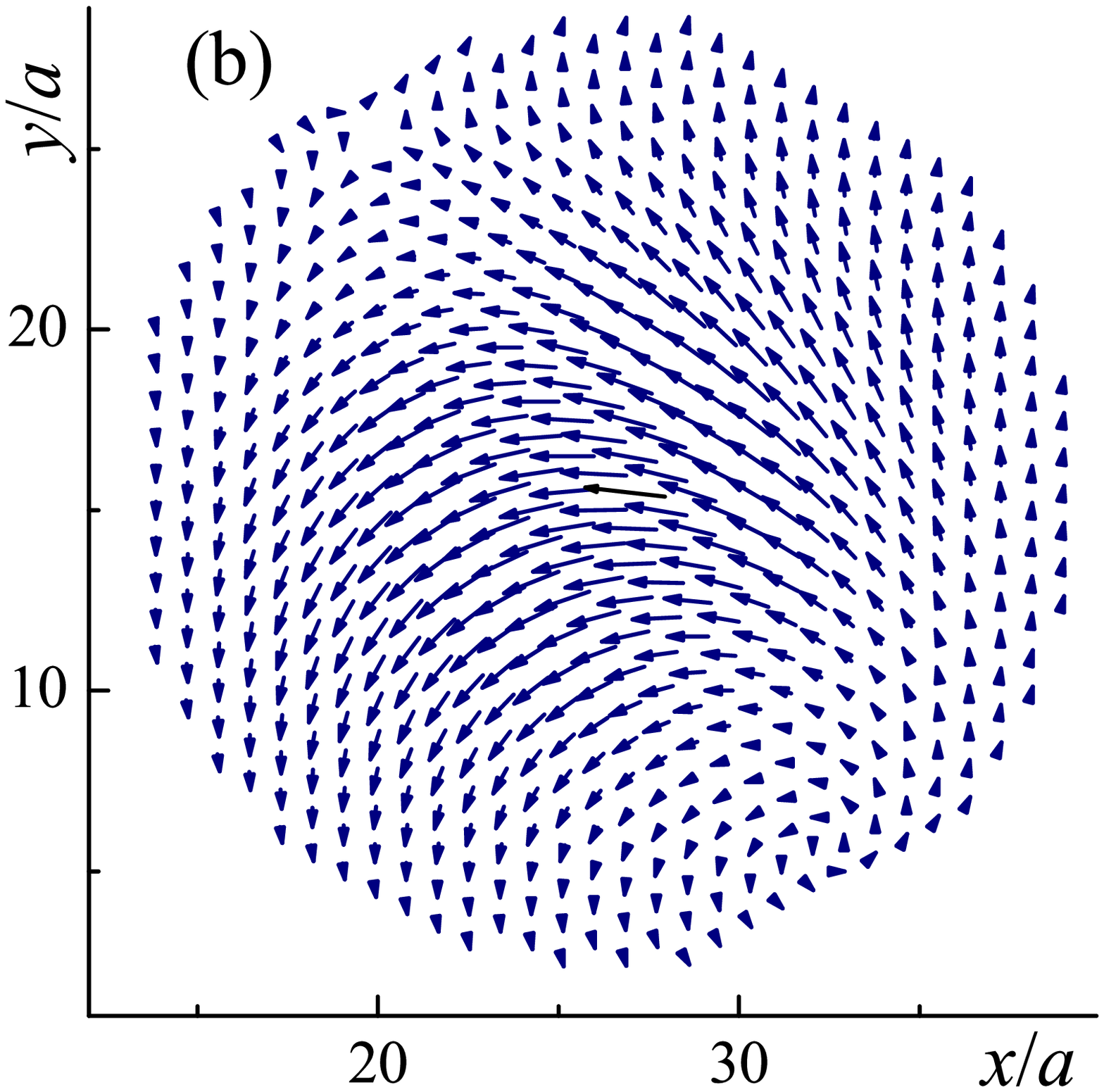}
\caption{\label{FigOn-siteSDWN-2Vec}The on-site spins
$\mathbf{S}_{\mathbf{n}1\alpha}$
in sublattice ${\cal A}$ (a) and
${\cal B}$ (b) at $x=-2$. Spins lie in the $xz$ plane. For the purpose of
the visual representation, the rotation of the spins from $xz$ plane to
$xy$ plane is performed. Only AA region of the supercell is shown.
}
\end{figure}

\begin{figure*}[t]
\centering
\includegraphics[width=0.3\textwidth]{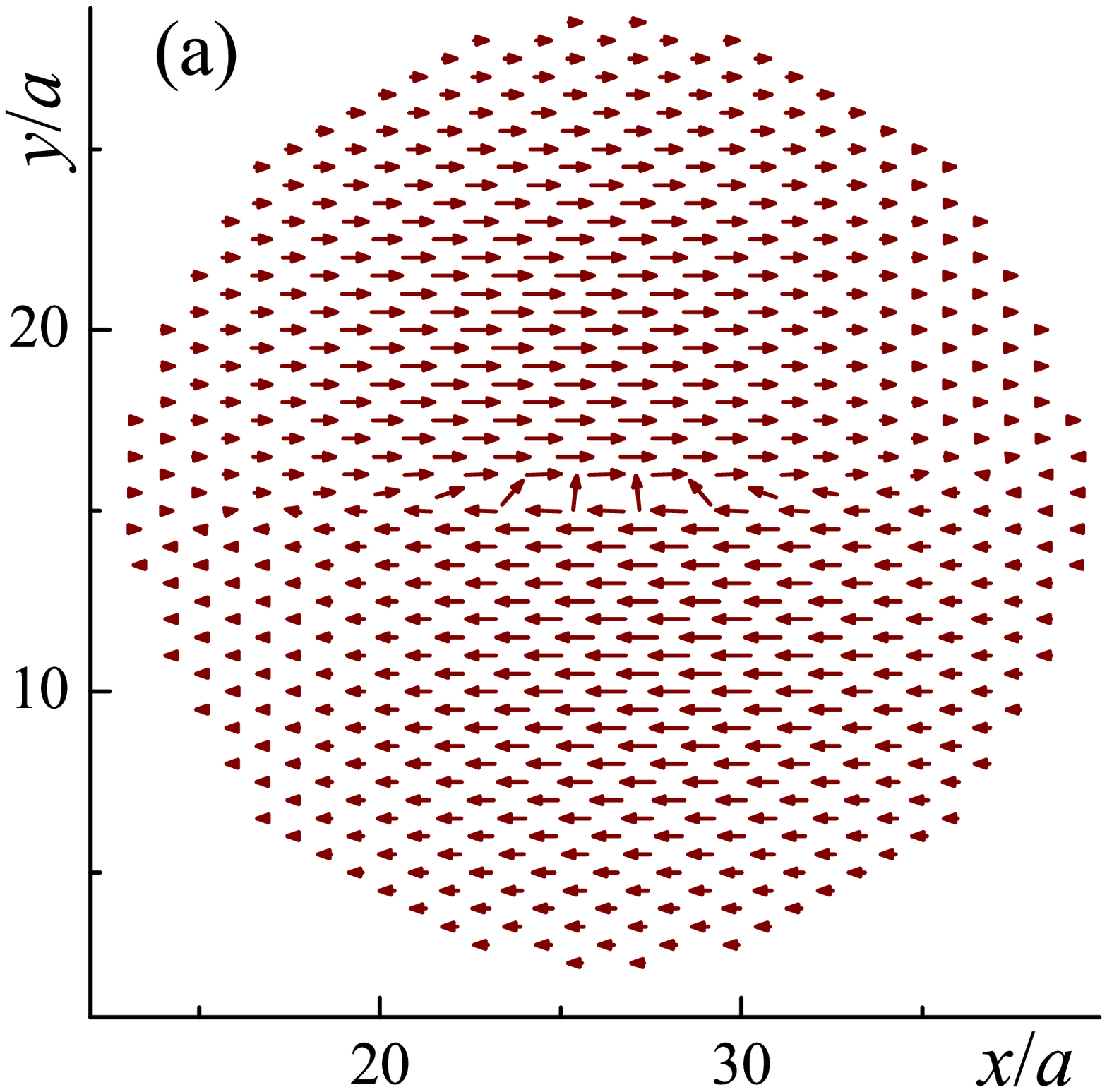}
\includegraphics[width=0.3\textwidth]{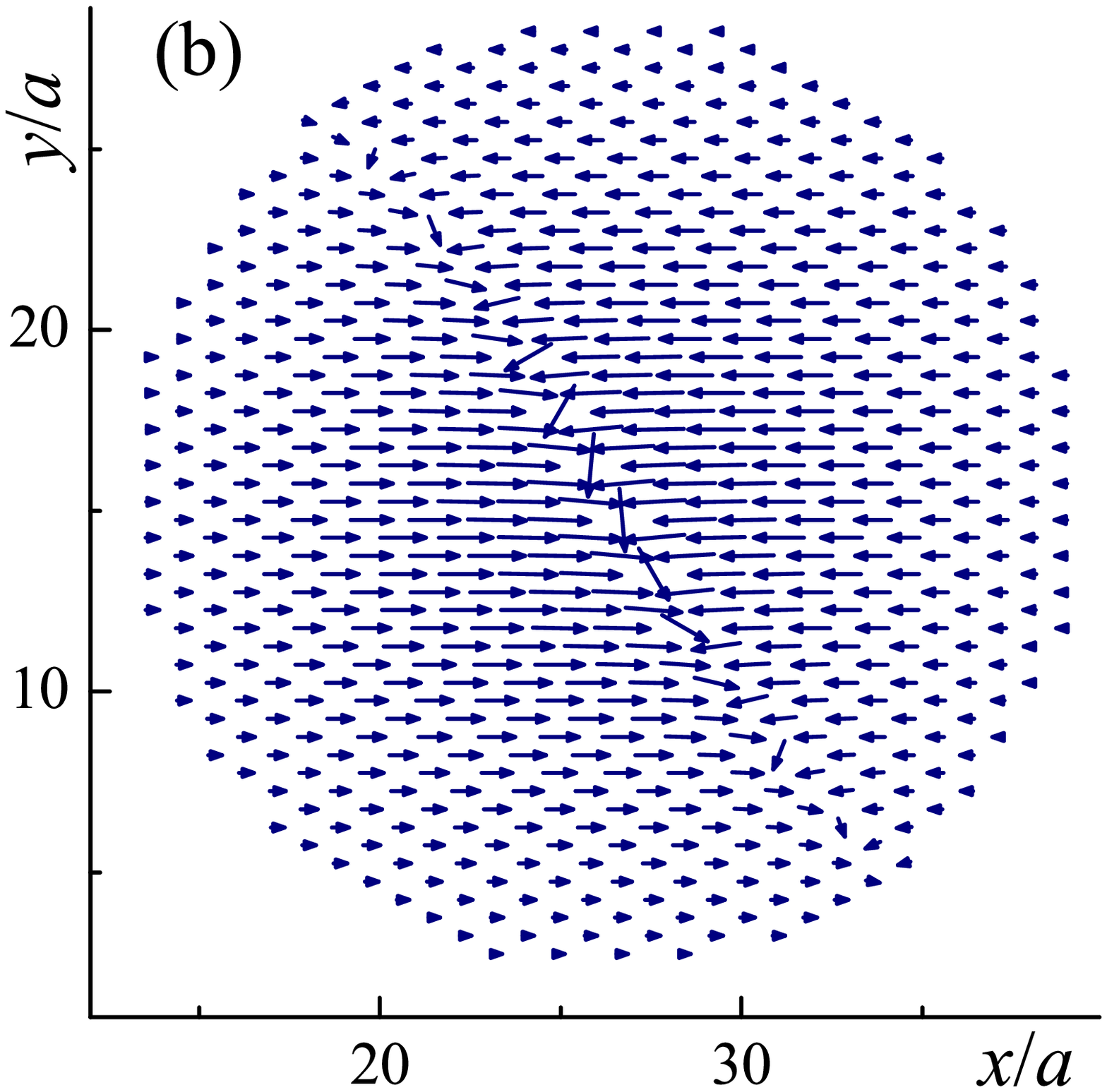}
\includegraphics[width=0.3\textwidth]{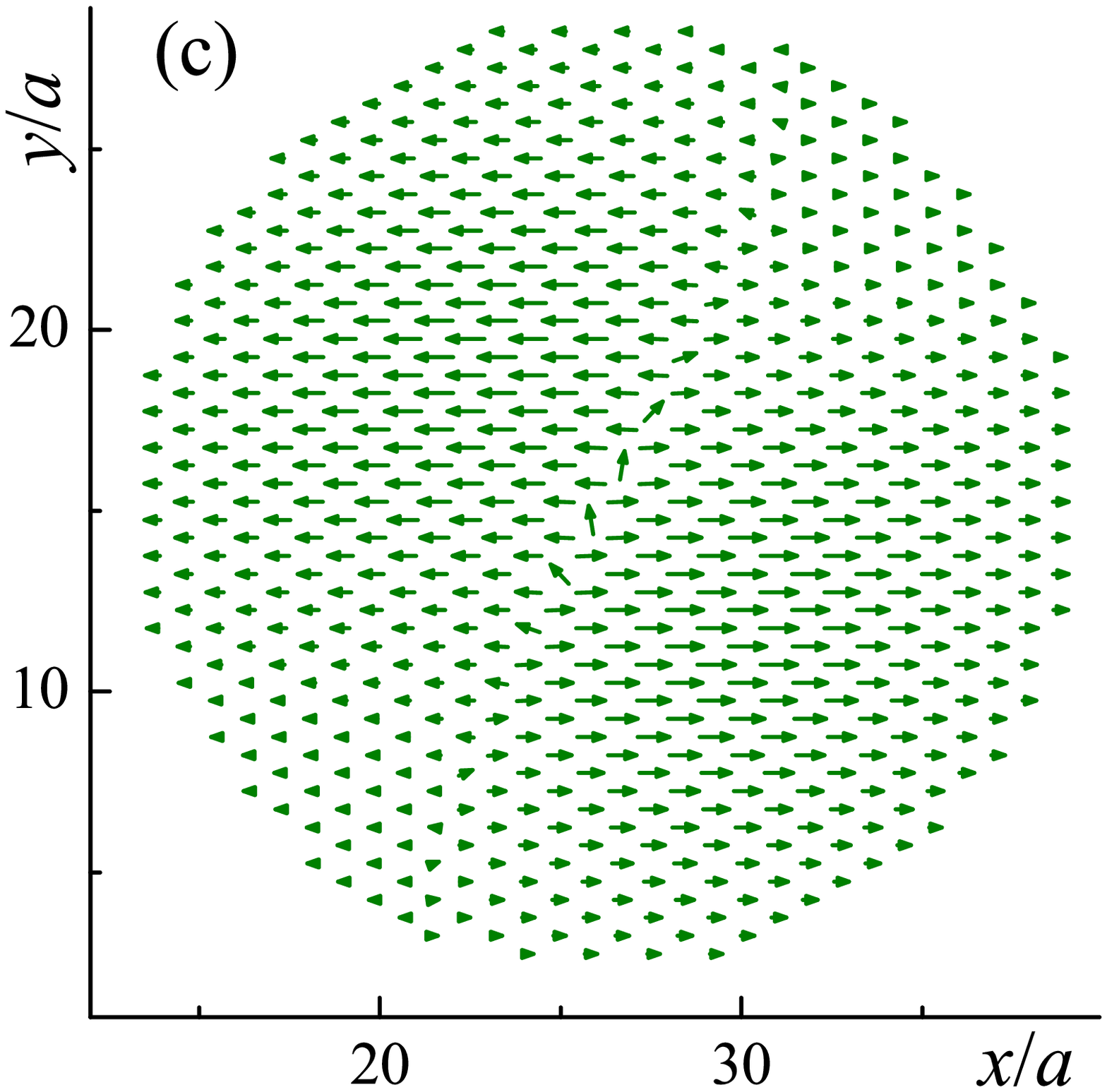}\\
\includegraphics[width=0.3\textwidth]{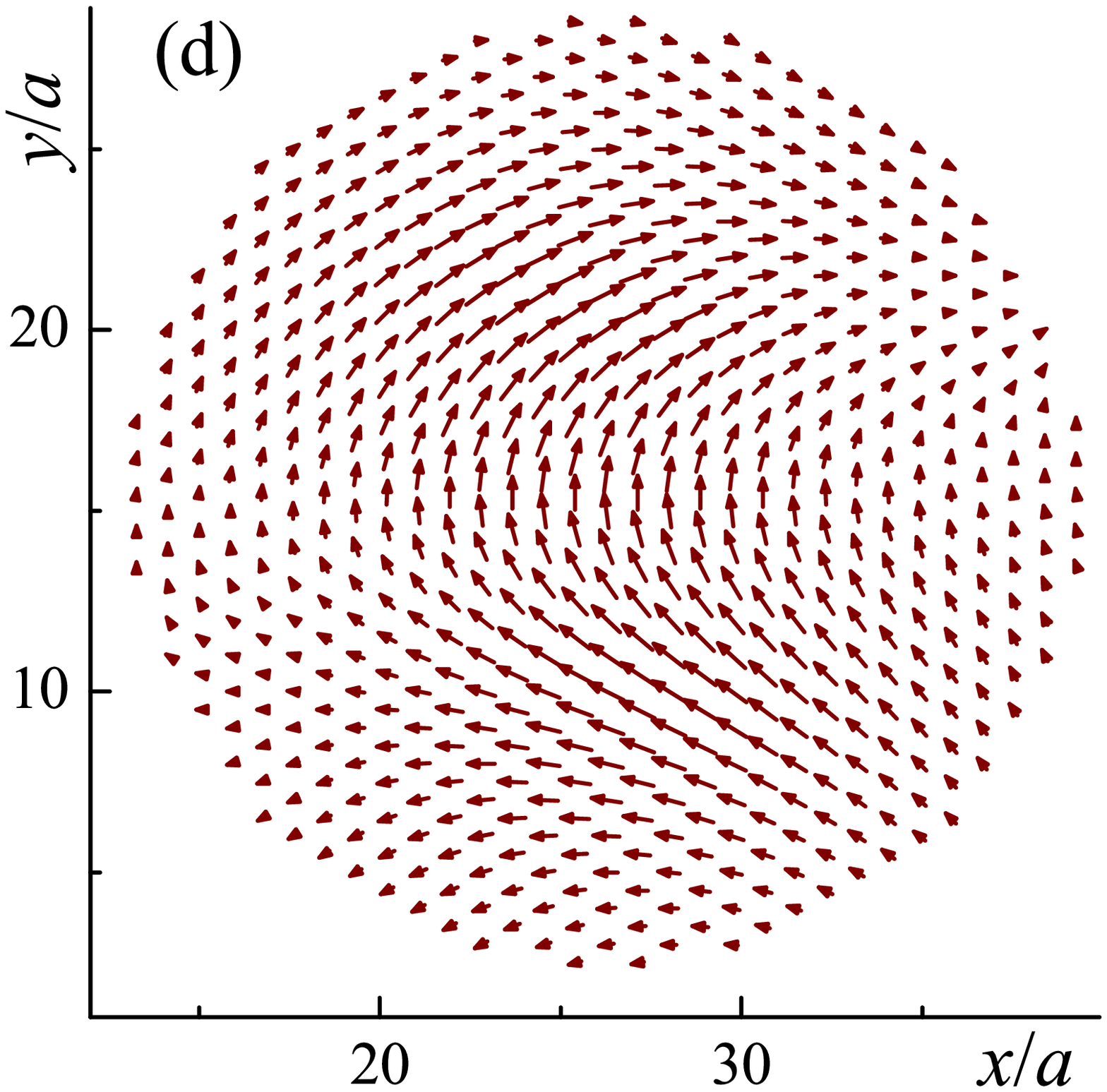}
\includegraphics[width=0.3\textwidth]{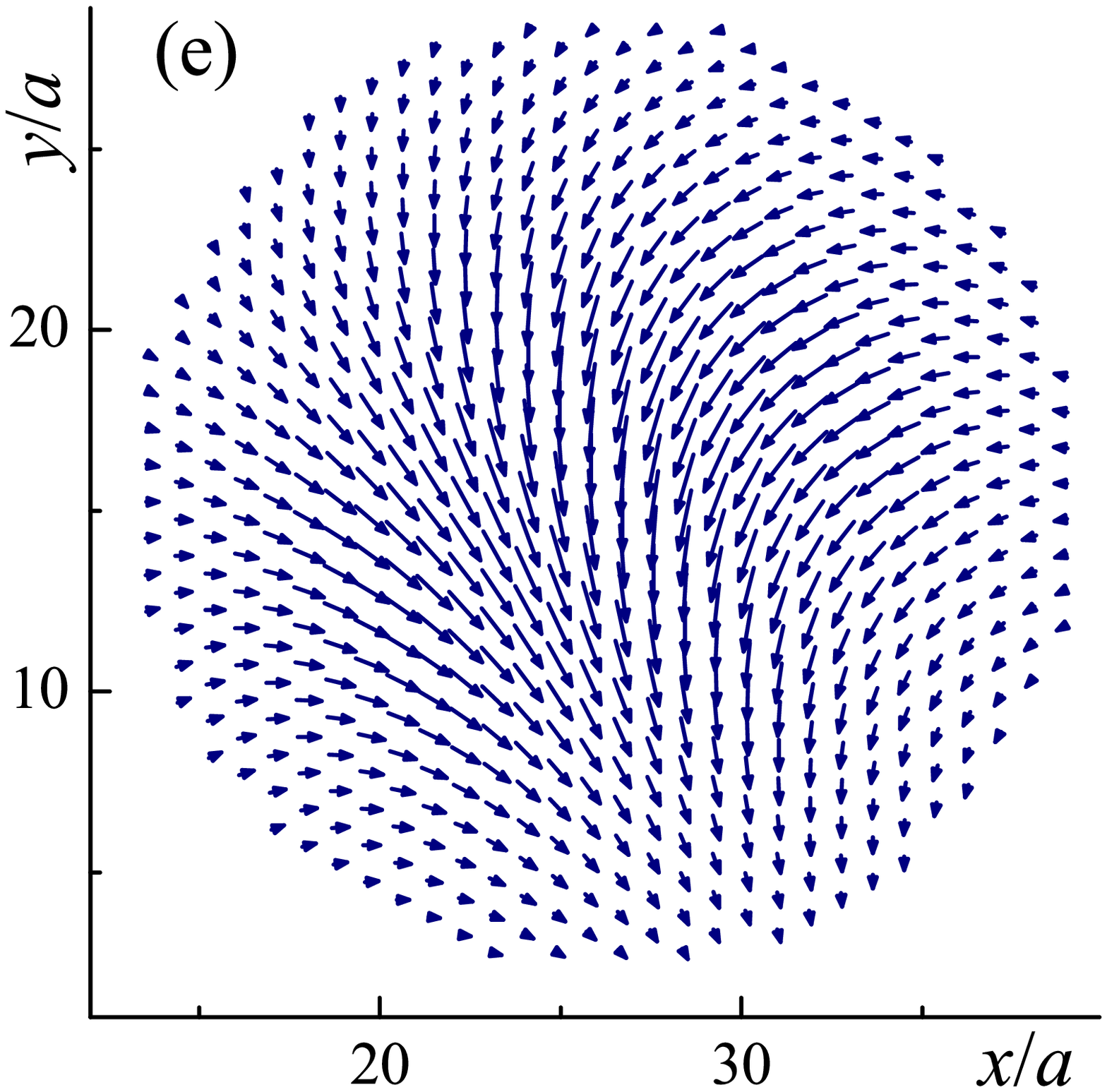}
\includegraphics[width=0.3\textwidth]{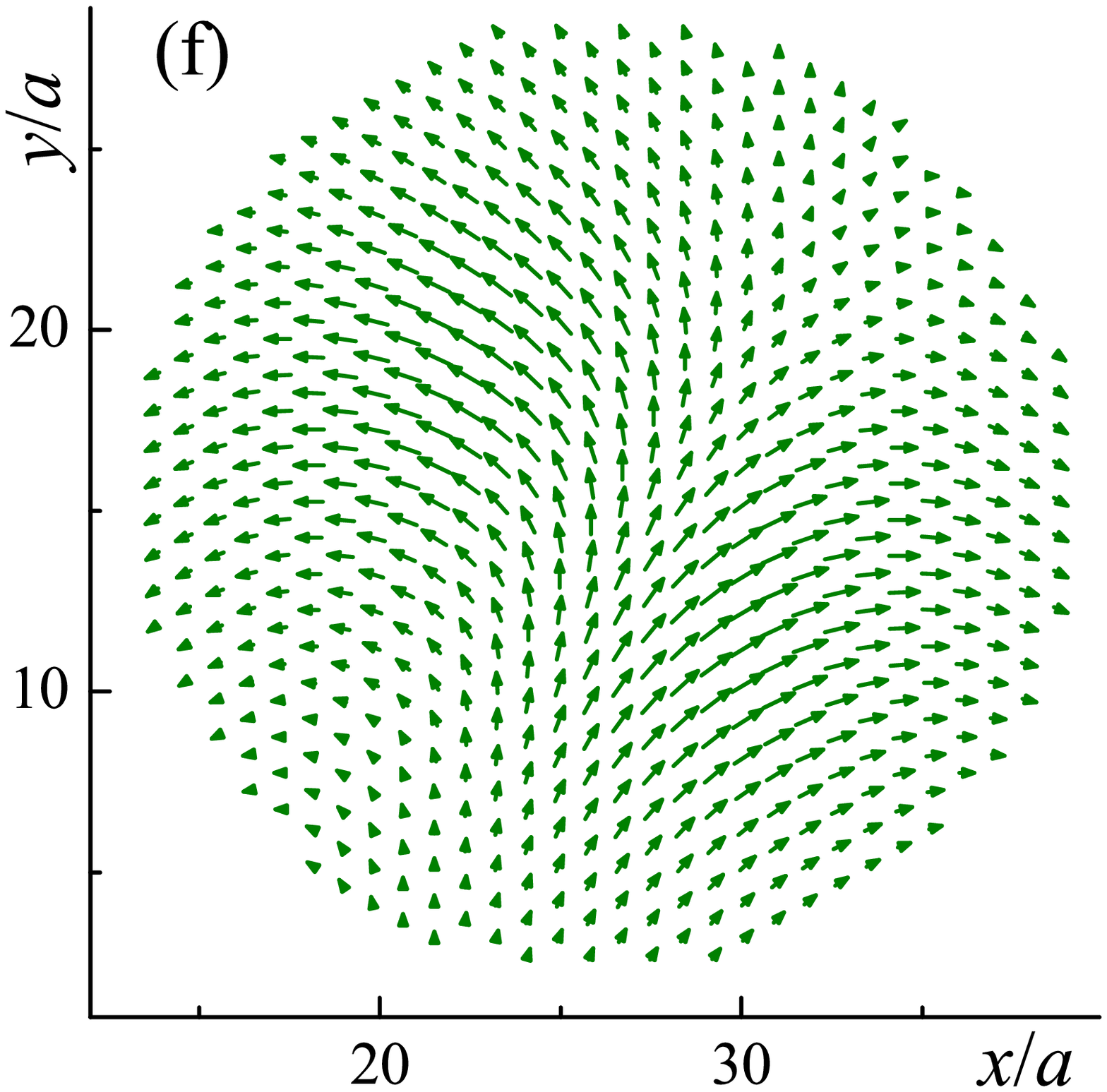}
\caption{\label{FigInter-siteSDWN-2Vec}
The textures of spins on the link,
$\mathbf{S}^{(\ell)}_{\mathbf{n}1}$,
calculated at $x=-2$. Panels (a)\,--\,(c) show the $xy$ projections of the
spins for different values of link parameter
$\ell=1,2,3$,
while panels (d)\,--\,(f) depicts the $xz$ projections (after rotation to
the $xy$ plane). Only AA region of the supercell is shown.
}
\end{figure*}

The contribution to $A^{c(\ell)}_{\mathbf{n}i}$ from excluded bands is non-zero
(moreover, one can check that it is non-zero even for	single-layer graphene). In the limit of uncoupled layers this contribution is independent of $\mathbf{n}$ and $i$. Thus, one can write $A^{c(\ell)}_{\mathbf{n}i}=\bar{A}+\delta\!A^{c(\ell)}_{\mathbf{n}i}$, where $\bar{A}$
can be estimated as
\begin{equation}
\bar{A}=\frac{V_{\text{nn}}}{2}\int\frac{d^2\mathbf{k}}{v_{BZ}}e^{-i\varphi_{\mathbf{k}}^{i}}\,\Theta\left(t|f_{\mathbf{k}}^{i}|-E_0\right),
\end{equation}
where $\varphi_{\mathbf{k}}^{i}=\arg(f_{\mathbf{k}}^{i})$, and $\bar{A}$ is
independent of $i$. For parameters chosen, we have
$\bar{A}/V_{\text{nn}}=0.26$. The parameter $\bar{A}$ renormalizes the
in-plane nearest-neighbor hopping amplitude $t$.
We assume that this renormalization is already absorbed into the value
$t=-2.57$\,eV,
and thus can be ignored.
As for $\delta\!A^{c(\ell)}_{\mathbf{n}i}$, we add quadratic term to the effective Hamiltonian, similar to that in Eq.~\eqref{Hrepl}.

To calculate the order parameters and the potentials we minimize the total
energy ${\cal E}$ iteratively using the method of successive
approximations. In each iteration, we calculate $E^{(S)}_{\mathbf{p}}$ and
$\Phi^{(S)}_{\mathbf{pG}i\alpha\sigma}$ of the matrix
$\hat{H}'_{\mathbf{p}}$, and the gradients
$\partial\cal E/\partial\lambda=\langle \partial H^{\text{MF}}/\partial\lambda\rangle$,
where
$\lambda=\Delta_{\mathbf{n}i\alpha}$,
$\Delta^{z}_{\mathbf{n}i\alpha}$,
$A^{(\ell)}_{\mathbf{n}i\sigma}$,
$A^{z(\ell)}_{\mathbf{n}i}$,
or $\delta\!A^{c(\ell)}_{\mathbf{n}i}$.
These gradients are used to calculate new values of the aforementioned
quantities. New value of
$\Delta^{c}_{\mathbf{n}i\alpha}$
is found using
Eq.~\eqref{Deltacnia_} with $\mu$ from Eq.~\eqref{x_}.

\textbf{Results.}
Our numerical calculations reveal that the SDW order exists in the system at any doping level. However, the spatial distributions of the SDW magnetization are qualitatively different for different $x$.

Let us start from $x=0$. Figures~\ref{FigSDW}(a)\,--\,(d) depicts the spatial distributions of
${\cal D}_{\mathbf{n}i\alpha}
=
U|\mathbf{S}_{\mathbf{n}i\alpha}|$
and
${\cal A}^{(\ell)}_{\mathbf{n}i}
=
V_{\text{nn}}|\mathbf{S}^{(\ell)}_{\mathbf{n}i}|$
inside the superlattice cell for
$i=1$.
These quantities describe the absolute values of the on-site spins and spins on the links, correspondingly. We see that the order parameters are non-zero only in the AA region. At zero doping, the order parameters spatial distributions have rather symmetric form. The area of non-zero ${\cal D}_{\mathbf{n}1\alpha}$ has a shape of a `rounded' hexagon. This hexagon is invariant under rotation on $60^{\circ}$ around the center of the AA region $\mathbf{R}_0=(\mathbf{R}_1+\mathbf{R}_2)/3$, with $\mathbf{R}_{1,2}$ being the elementary vectors of the superlattice. Figures~\ref{FigSDW}(b)\,--\,(d) demonstrate that spin magnetization on the links remain finite within the areas shaped like a dumbbell. The rotation on
$180^{\circ}$ and also on $120^{\circ}$ around $\mathbf{R}_0$ preserves ${\cal A}^{(\ell)}_{\mathbf{n}1}$.

When $x=0$ all spins, $\mathbf{S}_{\mathbf{n}i\alpha}$ and $\mathbf{S}^{(\ell)}_{\mathbf{n}i}$, are collinear. For layer $1$, all on-site spins $\mathbf{S}_{\mathbf{n}1{\cal A}}$ of the sublattice ${\cal A}$ point along $x$~axis, while in the sublattice ${\cal B}$ they have the opposite direction. The same is true for layer $2$. Therefore, the on-site spin texture has the collinear AFM arrangement. The spins on the links
$\mathbf{S}^{(\ell)}_{\mathbf{n}i}$
also form a kind of AFM structure: in one part of the dumbbell $\mathbf{S}^{(\ell)}_{\mathbf{n}i}$ directed along $x$~axis, while in another part of the dumbbell they have the opposite direction.

\begin{figure*}[t]
\centering
\includegraphics[width=0.22\textwidth]{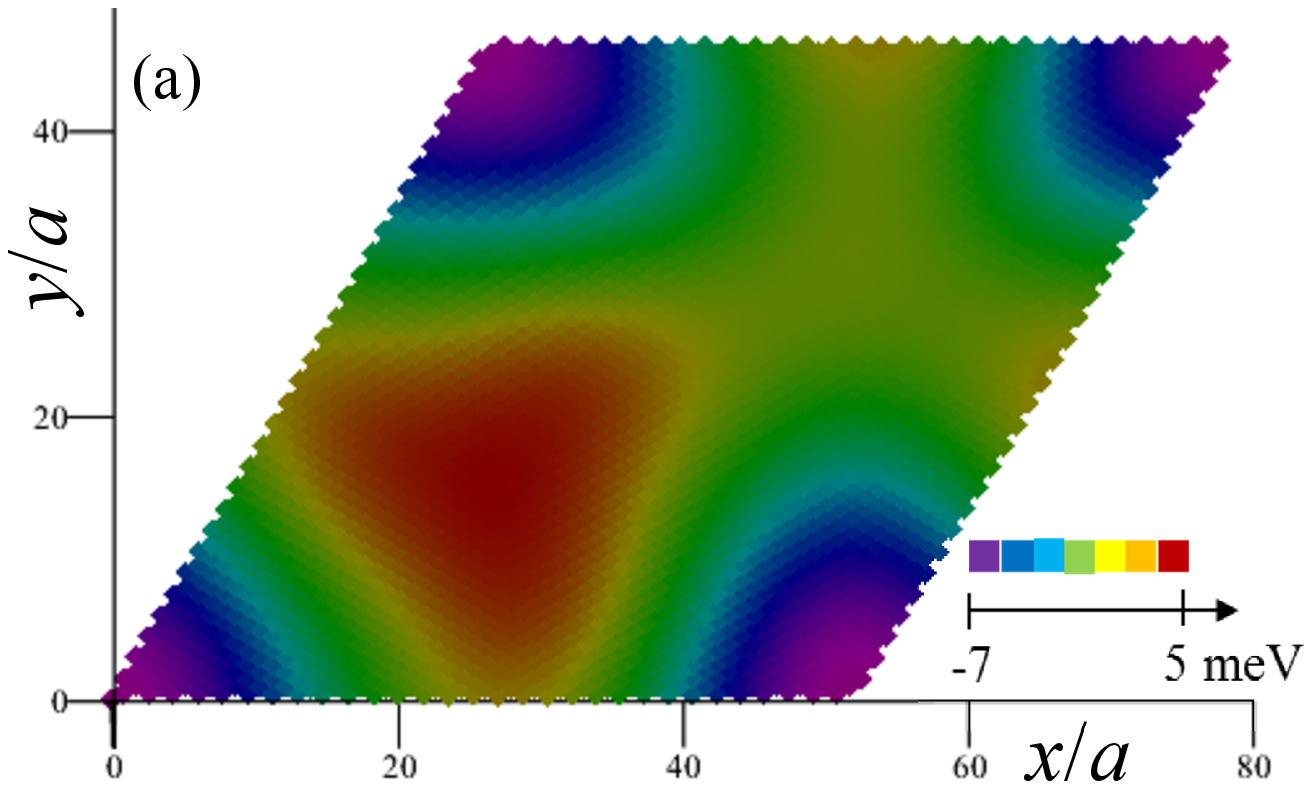}
\includegraphics[width=0.22\textwidth]{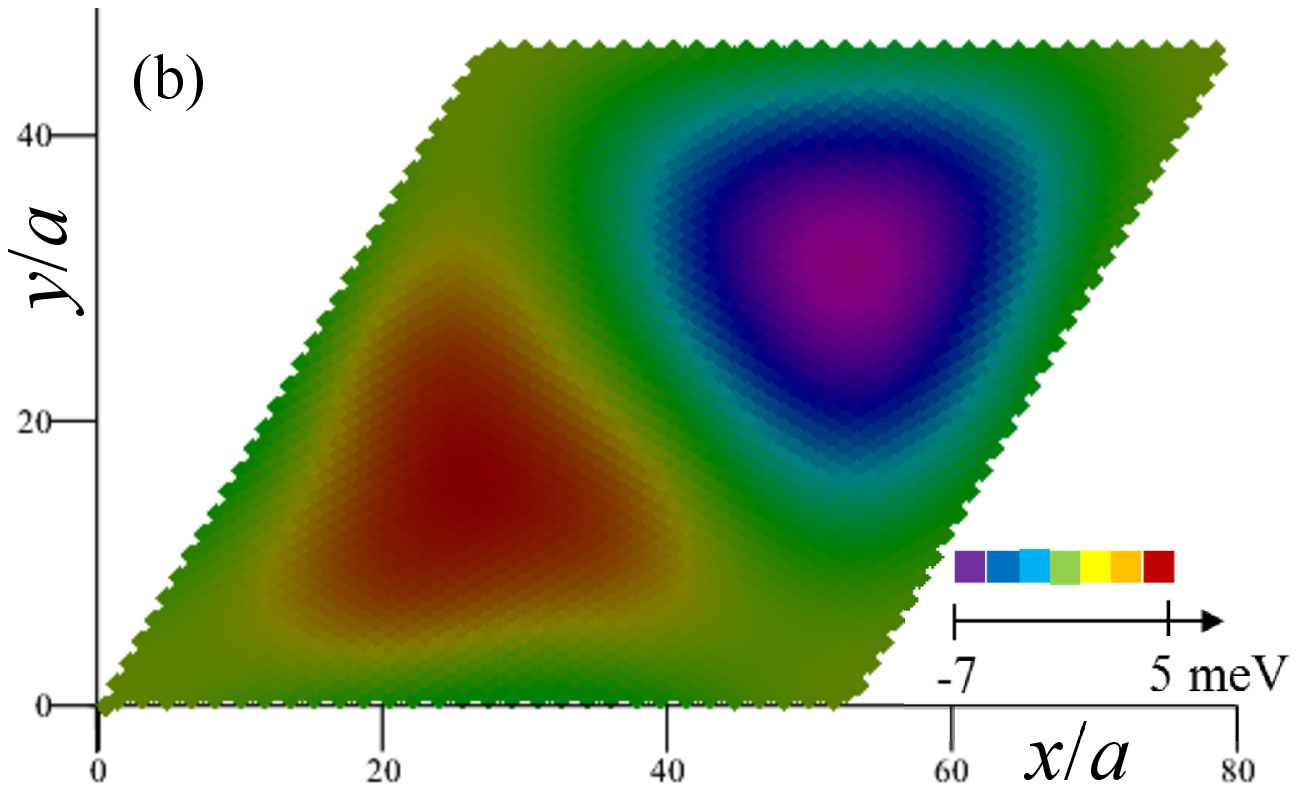}
\includegraphics[width=0.22\textwidth]{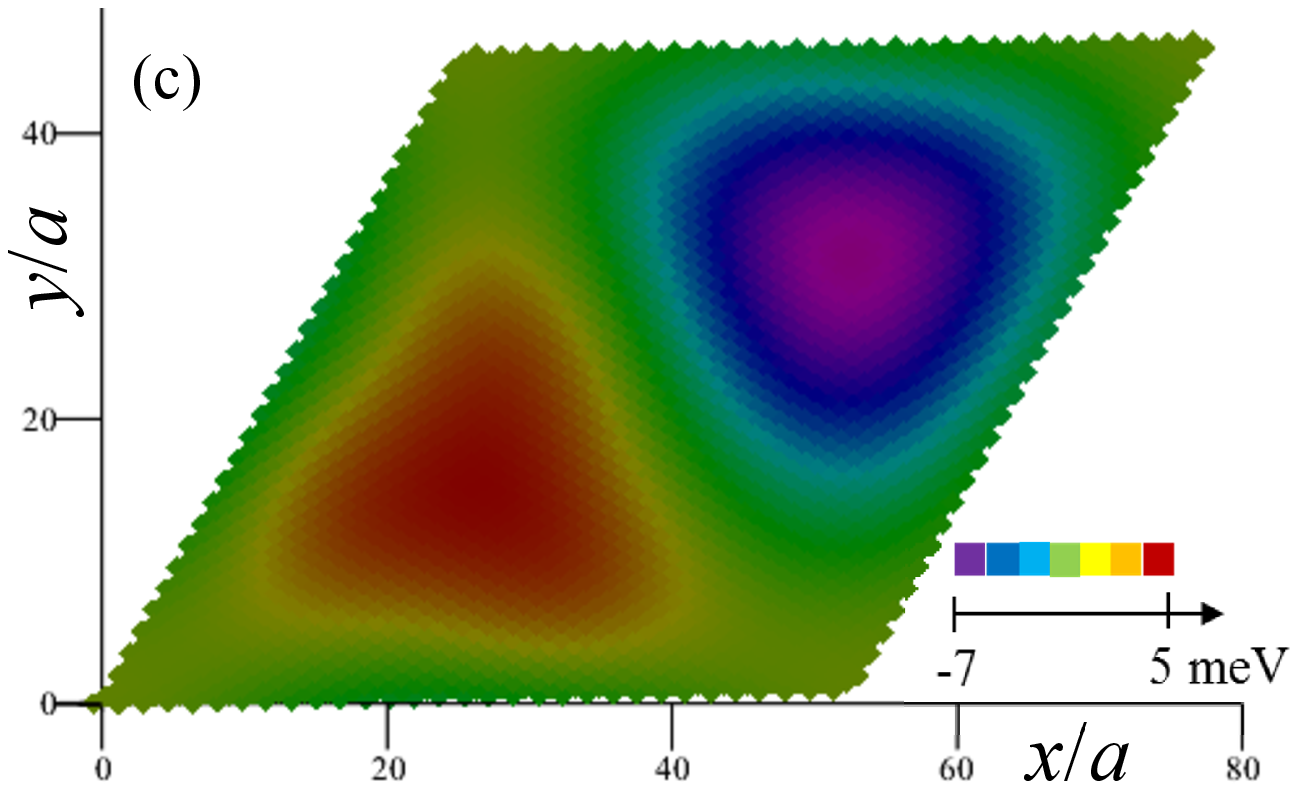}
\includegraphics[width=0.22\textwidth]{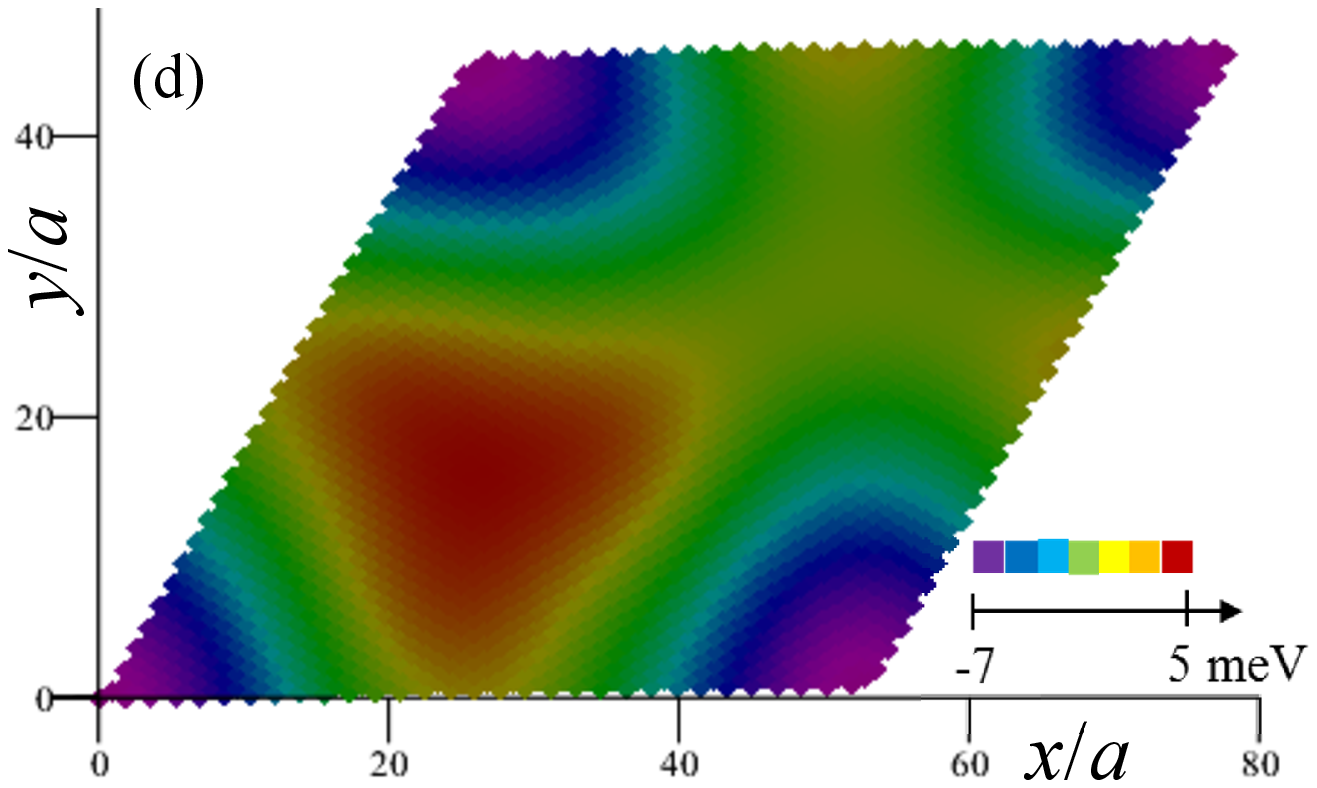}\\
\includegraphics[width=0.22\textwidth]{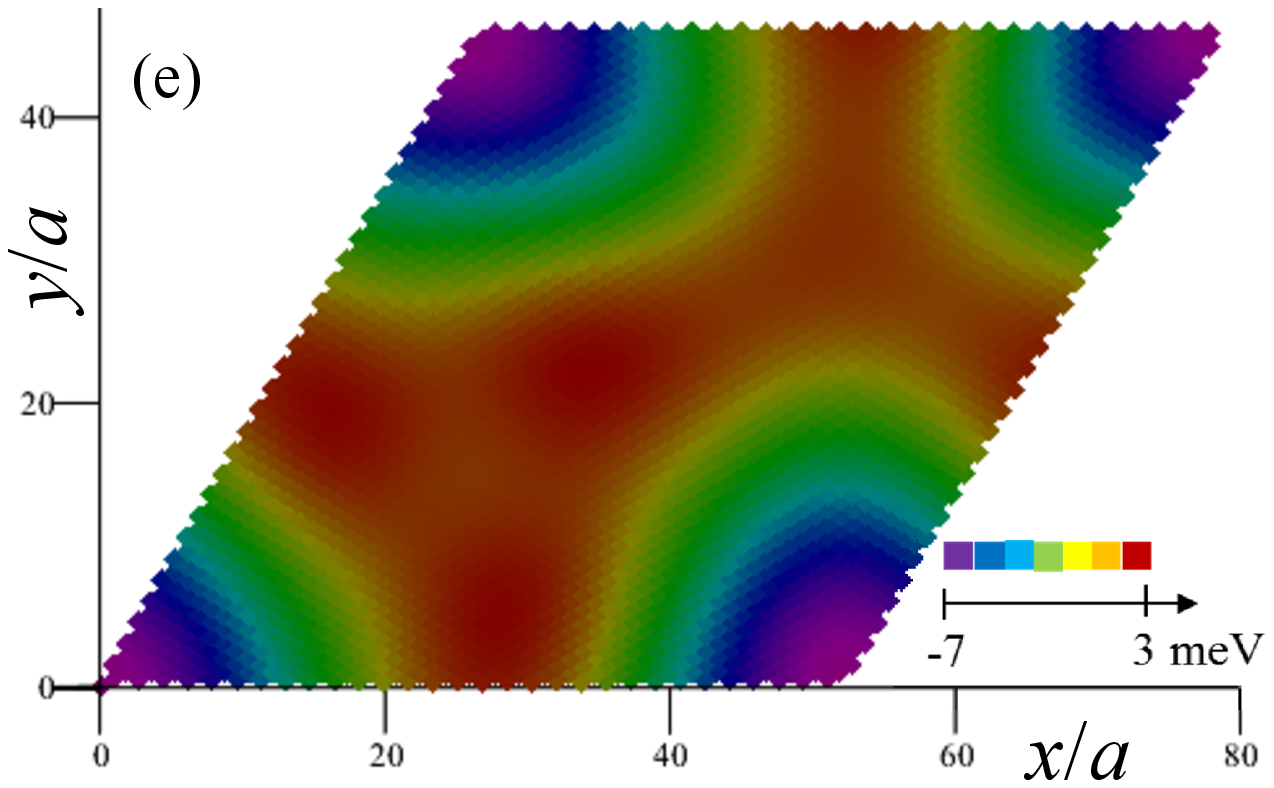}
\includegraphics[width=0.22\textwidth]{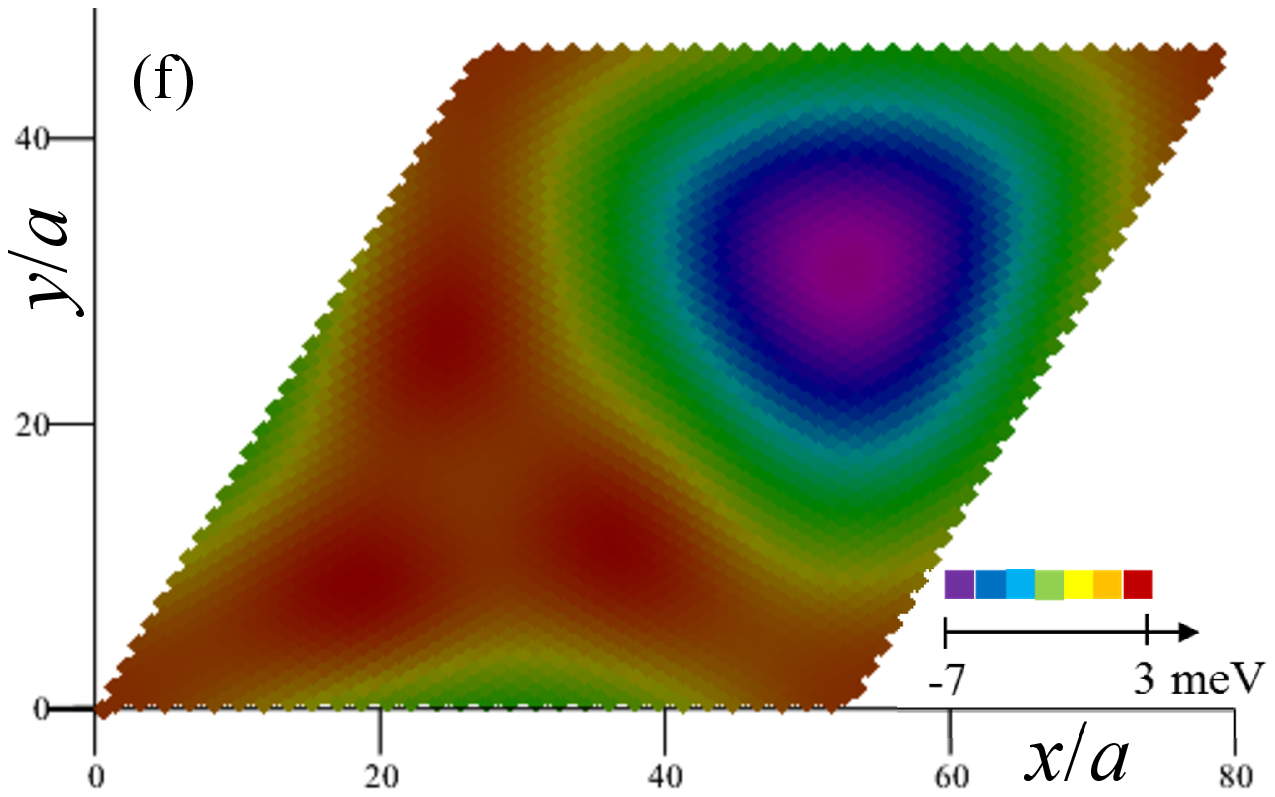}
\includegraphics[width=0.22\textwidth]{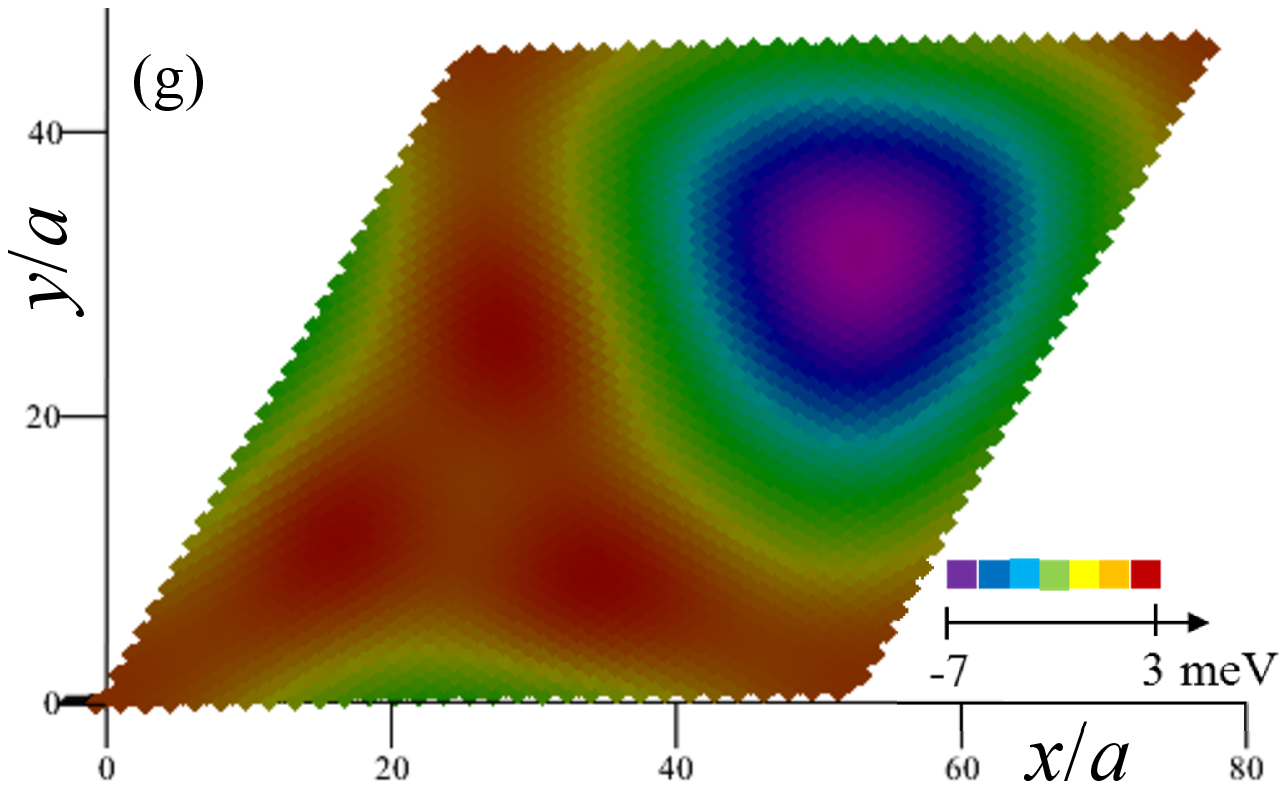}
\includegraphics[width=0.22\textwidth]{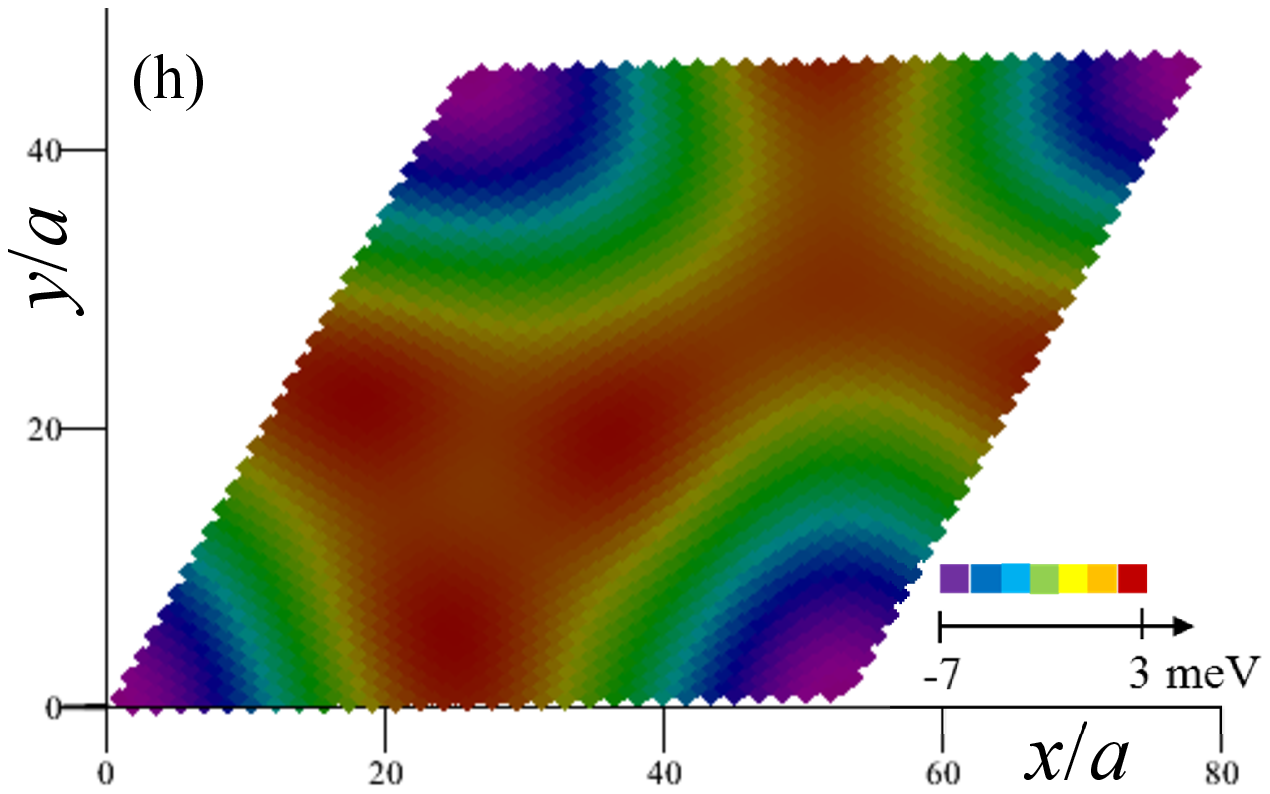}
\caption{\label{FigOn-siteCDW}
The spatial distribution of the on-site potential
$\Delta^c_{\mathbf{n}1{\cal A}}$
[(a), (e)],
$\Delta^c_{\mathbf{n}1{\cal B}}$
[(b), (f)],
$\Delta^c_{\mathbf{n}2{\cal A}}$
[(c), (g)], and
$\Delta^c_{\mathbf{n}2{\cal B}}$
[(d), (h)], calculated at $x=0$ [(a)\,--\,(d)] and at $x=-2$
[(e)\,--\,(h)].}
\end{figure*}

The order parameters distributions change with doping. In Fig.~\ref{FigSDW}(e)\,--\,(h) we plot the spatial distribution of ${\cal D}_{\mathbf{n}1\alpha}$ and ${\cal A}^{(\ell)}_{\mathbf{n}1}$ at hole doping $x=-2$. We see that now no hexagonal symmetry exists. The profile of ${\cal D}_{\mathbf{n}1\alpha}$ is stretched unidirectionally, and $120^{\circ}$ rotation is no longer a symmetry of ${\cal A}^{(\ell)}_{\mathbf{n}1}$. However, all these order parameters are symmetric under rotation on $180^{\circ}$. In doped systems the spin textures are no longer collinear. The on-site spins turn out to be coplanar. We believe that this is not an artifact since our numerical procedure accounts for non-coplanar textures. In our simulations the on-site spins $\mathbf{S}_{\mathbf{n}i\alpha}$ lie in the $xz$~plane (see Fig.~\ref{FigOn-siteSDWN-2Vec}), they  form a helical AFM structure. The spins on links, $\mathbf{S}^{(\ell)}_{\mathbf{n}i}$, form a non-coplanar structures. However, almost all of these spins lie in the $xz$~plane, and only small fraction of them have $y$ components. Figures~\ref{FigInter-siteSDWN-2Vec}(a)\,--\,(c) show the directions of spins $\mathbf{S}^{(\ell)}_{\mathbf{n}i}$ projected to the $xy$~plane, while Figs.~\ref{FigInter-siteSDWN-2Vec}(d)\,--\,(f) show the directions of spins projected to the $xz$~plane (with subsequent rotation to the $xy$~plane). We see that almost all spins lie in the $xz$~plane. The spins which violate the coplanarity lie along three lines passing through the center of the AA~region. These lines are visible in Fig.~\ref{FigInter-siteSDWN-2Vec}(a)\,--\,(c).

\begin{figure*}[t]
\centering
\includegraphics[width=0.3\textwidth]{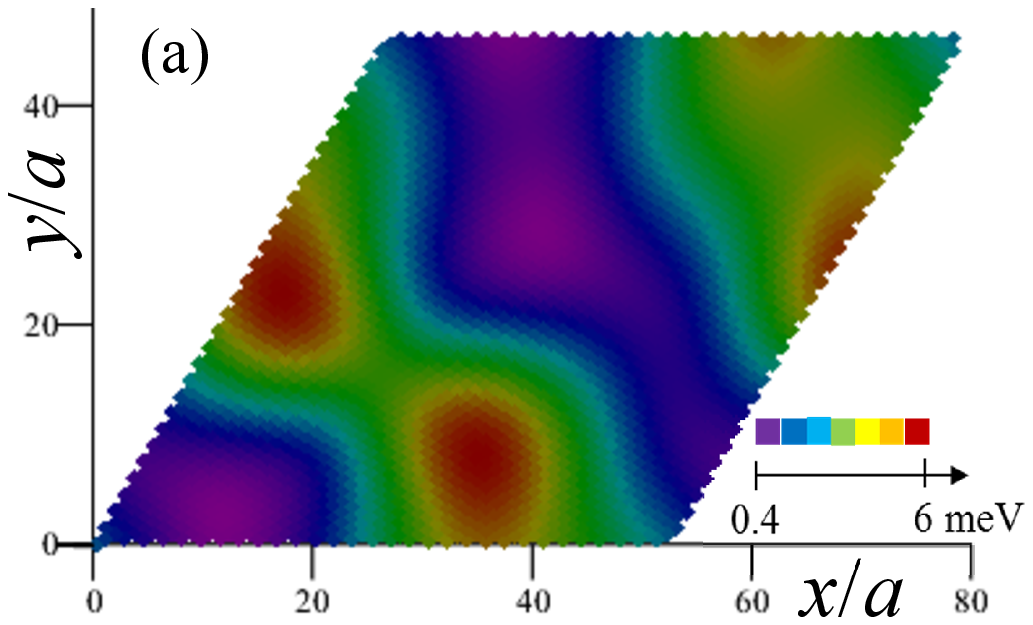}
\includegraphics[width=0.3\textwidth]{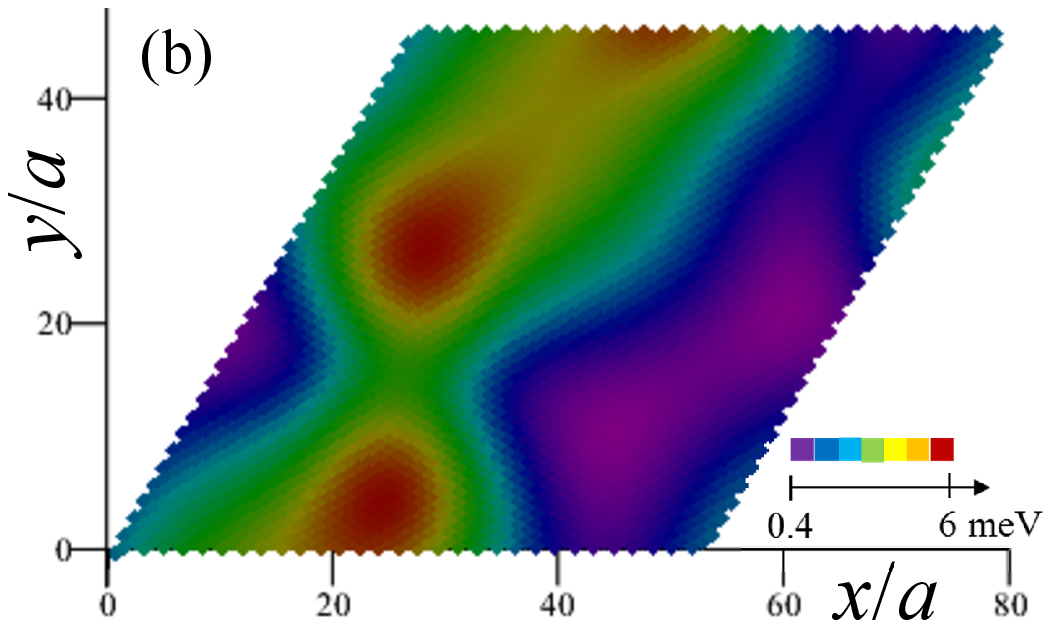}
\includegraphics[width=0.3\textwidth]{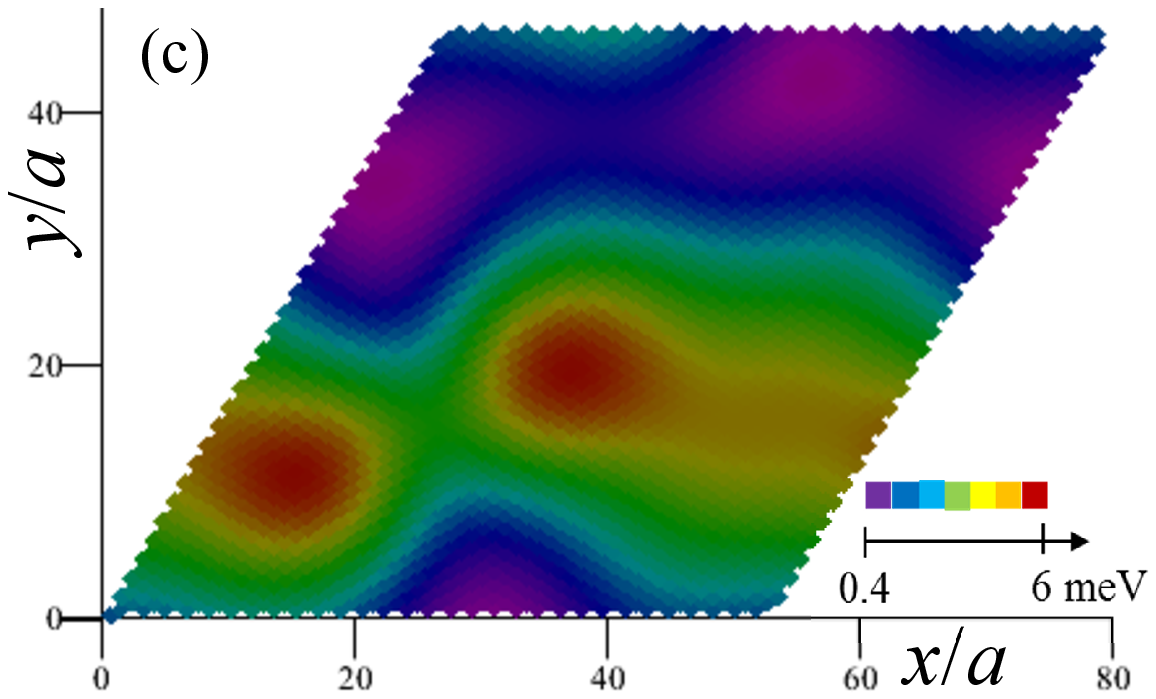}\\
\includegraphics[width=0.3\textwidth]{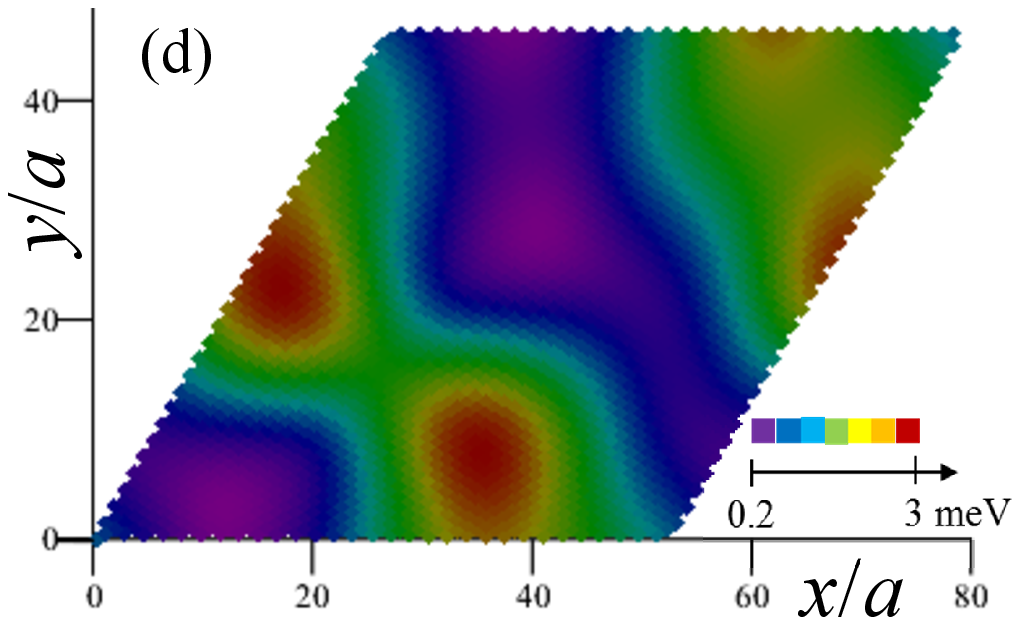}
\includegraphics[width=0.3\textwidth]{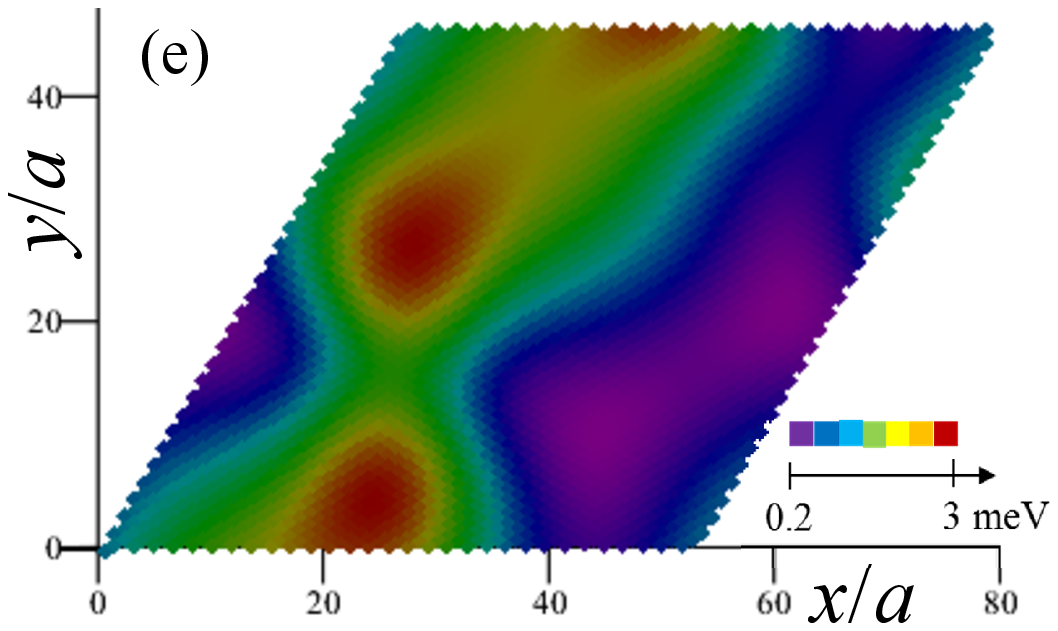}
\includegraphics[width=0.3\textwidth]{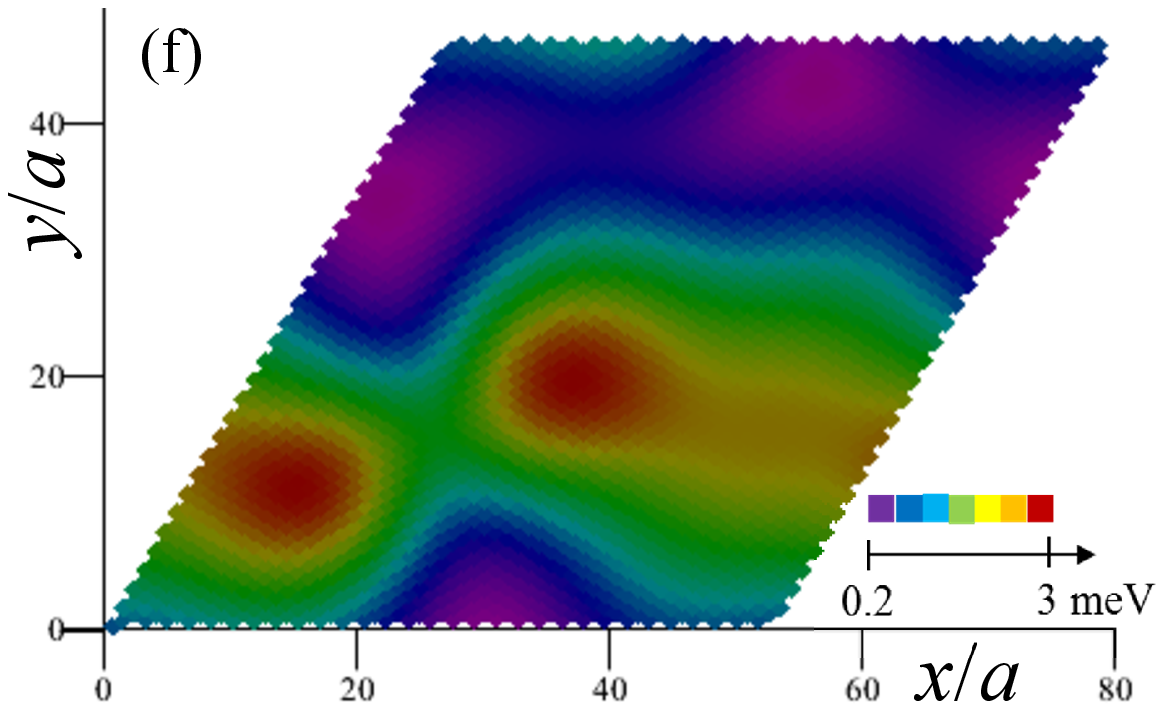}
\caption{\label{FigInter-siteCDW}
The spatial distribution of the inter-site potentials $\delta A^{c(\ell)}_{\mathbf{n}1}$ for three different $\ell$ , calculated at x=0 [(a)\,--\,(c)] and at $x=-2$ [(d)\,--\,(f)].}
\end{figure*}

In Fig.~\ref{FigOn-siteCDW} we plot the spatial distributions of the on-site potential $\Delta^c_{\mathbf{n}i\alpha}$ calculated at zero doping [panels (a)\,--\,(d)] and at $x=-2$ [panels (e)\,--\,(h)]. The distributions are shown for each layer and sublattice separately. At $x=0$ there is an excess of electrons in the AA region and lack of electrons in the AB (corners of the supercell) and BA (area centered at $\mathbf{R}_{BA}=2(\mathbf{R}_1+\mathbf{R}_2)/3$) regions of the superlattice cell. At $x=-2$, the area, where excess of electrons is observed, expands; extra electrons also appear in the AB (for $\Delta^c_{\mathbf{n}1{\cal B}}$ and $\Delta^c_{\mathbf{n}2{\cal A}}$) and BA (for $\Delta^c_{\mathbf{n}1{\cal A}}$ and $\Delta^c_{\mathbf{n}2{\cal B}}$) regions. At zero doping the charge distributions remain invariant under rotation on $120^{\circ}$ around the point $\mathbf{R}_0$. The potentials $\Delta^c_{\mathbf{n}i{\cal A}}$ and $\Delta^c_{\mathbf{n}i{\cal B}}$ are transformed into each other under the rotation on $60^{\circ}$ around the point $\mathbf{R}_0$. The same is true approximately at $x=-2$. Thus, the nematicity does not manifest itself through the charge distribution.

Figure~\ref{FigInter-siteCDW} shows the spatial distributions of the inter-site potentials $\delta A^{c(\ell)}_{\mathbf{n}1}$, calculated at $x=0$ and $x=-2$. The profiles of $\delta A^{c(\ell)}_{\mathbf{n}1}$ are approximately the same for $x=0$ and $x=-2$, and only absolute values of $\delta A^{c(\ell)}_{\mathbf{n}1}$ change. These spatial distributions are approximately symmetric under rotation on $180^\circ$ around $\mathbf{R}_0$. The inter-site potentials are transformed into each other under the rotation on $120^\circ$ around the point $\mathbf{R}_0$.

\begin{figure}[t]
\centering
\includegraphics[width=0.45\textwidth]{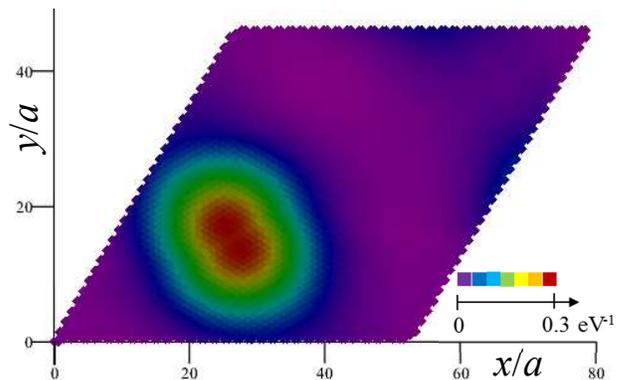}
\caption{\label{FigLDOS}The local density of states calculated close to half filling ($x=-1.75$). The spatial profile is stretched indicating the appearance of the electron nematic state.}
\end{figure}

Thus, the doping reduces the symmetry of the SDW order from
$C_6$,
which is the symmetry of the lattice, to
$C_2$
indicating the appearance of the nematic
state~\cite{NematicPRB2020}.
At the same time the symmetry of the charge-related quantities
$\Delta^c$
and
$A^{c(\ell)}$
is virtually unaffected by doping. These parameters are integrated
quantities that are influenced by contributions of high-energy states, the
latter being insensitive to the low-energy symmetry breaking. At the same
time, the electron nematic state can manifest itself as a symmetry
reduction of the local density of states, which can be detected in an STM
experiment. This is confirmed by our simulations.
Figure~\ref{FigLDOS}
shows the spatial distribution of the local density of states calculated
close to half filling (at
$x=-2$
the density of states is almost zero). The spatial profile is stretched
indicating the appearance of the nematic state. Such a feature was observed
in the
experiments~\cite{MottNematicNature2019,KerelskyNematicNature2019,
STMNature2019,WongNematic2020}.

Finally, we compare the values of the spin and charge density variations. At zero doping, the on-site spin order parameter ${\cal D}_{\mathbf{n}i\alpha}$ reaches the maximum about $34$\,meV. At the same time, the on-site charge density variation $\Delta^{c}_{\mathbf{n}i\alpha}$ is about $6$ times smaller. The SDW order parameters decrease down to zero, when the doping changes from $x=0$ to $x=\pm4$. At the same time, the values of $\Delta^c$ do not change much with $x$. However, even at $x=\pm2$ the maximum value of ${\cal D}_{\mathbf{n}i\alpha}$ is about $3$ times larger than that of $|\Delta^{c}_{\mathbf{n}i\alpha}|$, and only at $|x|\approx3$, the on-site SDW order parameter becomes comparable to $|\Delta^{c}_{\mathbf{n}i\alpha}|$.

\textbf{Conclusions.}
We study numerically the coexisting spin and charge density waves in the magic-angle tBLG in the doping range  $x=\pm 4$ extra electrons per supercell using the mean-field approach. The single-electron spectrum of the material has 8 almost flat almost degenerate bands. The electron-electron interaction breaks down the symmetries of the single-particle state forming a set of order parameters. We calculate self-consistently the charge distribution in the supercell and spin structure of the SDW order parameters. We found that the SDW order is stable in the whole doping range. The spin texture of the SDW order parameters depends crucially on $x$ changing from collinear at $x=0$ to almost coplanar at finite doping. The ground state of the doped system has the nematic symmetry.

We are grateful to the Joint Supercomputer Center of the Russian Academy of Sciences (JSCC RAS) for the computational resources provided. The data analysis and analytical calculations were funded by the Russian Science Foundation (project No.~22-22-00464
\url{https://rscf.ru/en/project/22-22-00464/}).



\end{document}